\documentclass[aps,groupedaddress,nofootinbib,floatfix,epfs,preprintnumbers]{revtex4}
\usepackage{amsfonts,amscd,amsmath,amssymb,graphicx,color,float}
\usepackage[section]{placeins}
\def\ep{\text{e}}
\def\g{\mathsf{g}}
\def\oh{\frac{1}{2}}
\def\s{\mathsf{s}}

\def\k{\mathsf{k}}
\def\n{\mathsf{n}}
\def\qqb{\text{\tiny q}\bar{\text{\tiny q}}}
\def\QQq{\text{\tiny QQq}}
\def\Qqq{\text{\tiny Qqq}}
\def\Q2q3{\text{\tiny QQqq}\bar{\text{\tiny q}}}
\def\QQqq{\text{\tiny QQ}\bar{\text{\tiny q}} \bar{\text{\tiny q}}}
\def\Qqbqq{\bar{\text{\tiny Q}}\text{\tiny qq}\bar{\text{\tiny q}}}
\def\QQ{\text{\tiny QQ}}
\def\Qq{\text{\tiny Qq}}
\def\nucl{\text{\tiny qqq}}

\def\qs{{\text q}}
\def\vs{{\text v}}

\def\rqqq{r_{3q}}
\def\qsn{{\text q}_3}

\def\km{-\frac{1}{4}\ep^{\frac{1}{4}}}

\def\Vz{{\text v}_{\text{\tiny 0}}}
\def\Vo{{\text v}_{\text{\tiny 1}}}

\def\rq{r_q}

\def\rqb{r_{\bar q}}
\def\rv{r_v}
\def\r0{r_0}
\def\rvb{r_{\bar v}}
\def\rqb{r_{\bar q}}

\def\2Qq{\text{\tiny Q}\bar{\text{\tiny Q}}{\text{\tiny q}}\bar{\text{\tiny q}}}

\def\Qqqq{\text{\tiny Qq}\bar{\text{\tiny q}}\bar{\text{\tiny q}}}
\def\Qqb{\text{\tiny Q}\bar{\text{\tiny q}}}

\def\QQb{\text{\tiny Q}\bar{\text{\tiny Q}}}

\def\3Q{3\text{\tiny Q}}
\def\qQb{\text{\tiny q}\bar{\text{\tiny Q}}}

\begin{document}
\preprint{LMU-ASC 26/22}
\title{$QQqq\bar q$ Quark System, Compact Pentaquark, and Gauge/String Duality}
\author{Oleg Andreev}
\thanks{Also on leave from L.D. Landau Institute for Theoretical Physics}
\affiliation{Arnold Sommerfeld Center for Theoretical Physics, LMU-M\"unchen, Theresienstrasse 37, 80333 M\"unchen, Germany}
\begin{abstract} 
For the case of two light flavors we propose the stringy description of the system consisting of two heavy and three light quarks, with the aim of exploring the quark organization inside the system and its low-lying Born-Oppenheimer potentials as a function of the heavy quark separation. Our analysis reveals several critical separations related to the processes of string reconnection, breaking and junction annihilation. We find that a compact pentaquark configuration makes the dominant contribution to the potential of the first excited state at small separations, and for separations larger than $0.1\,\text{fm}$, the antiquark-diquark-diquark structure emerges. Moreover, it turns out that the length scale of string junction annihilation is the same as that for the $QQ\bar q\bar q$ system. We also discuss the relation to the potential of the $QQq$ system and some relations among the masses of hadrons in the heavy quark limit. 
\end{abstract}
\maketitle

\vspace{0.85cm}
\section{Introduction}
\renewcommand{\theequation}{1.\arabic{equation}}
\setcounter{equation}{0}

Since the first observation \cite{X38} of the charmonium-like peak $X(3872)$, more than $60$ new hadrons have been observed at high statistical significance. Most of those are potentially exotic states such as tetraquarks, pentaquarks, hybrid mesons, and glueballs.\footnote{For more details, see the recent review article \cite{lebed} and the book \cite{book}.} More recently, sightings of pentaquarks consisting of three light quarks and one $c\bar c$ pair have been reported \cite{P43}. These are the example of $Q\bar Qqqq$ states. In general, there could exist other doubly heavy pentaquark states which are of type $QQqq\bar q$. But so far they have not been observed experimentally. 

One way to deal with doubly heavy quark systems is as follows. Because of the large ratio of the quark masses, the Born-Oppenheimer (B-O) approximation borrowed from atomic and molecular physics \cite{bo} seems justified.\footnote{For the further development of these ideas in the context of QCD, see \cite{braat}.} In that case the corresponding B-O potentials are defined as the energies of stationary configurations of the gluon and light quark fields in the presence of the static heavy quark sources. The hadron spectrum is then calculated by solving the Schr\"odinger equation in these potentials. 

Although lattice gauge theory is a well-established approach to non-perturbative QCD, it still remains to be seen what it can and can't do with regard to the doubly heavy pentaquark systems. Meanwhile, the gauge/string duality is a powerful way to understand gauge theories outside of the weak coupling limit, and therefore may be used as an alternative approach to gain important physical insights into this problem.\footnote{The book \cite{uaw} gives a thorough review of much of what is known about the gauge/string duality in relation to QCD.} Also within this approach, matters relating to pentaquarks have not been discussed in the literature. Filling the latter gap is one of the aims of the present paper.

The paper continues our study \cite{a-QQq, a-QQqq, a-QQbqqb} on the doubly heavy quark systems. It is organized as follows. We begin in Sec.II by recalling some preliminary results and setting the framework. Then in Sec.III, we construct and analyze a set of string configurations in five dimensions which provide a dual description of the $QQ qq\bar q$ system in the heavy quark limit. Among those we find the configurations relevant to the two low-lying B-O potentials. In the process, we introduce several length scales. These characterize transitions between different dominant configurations, and in fact, are related to different types of string interactions: reconnection, breaking and junction (baryon vertex) annihilation. We go on in Sec.IV to discuss further improvements and extensions. Finally, in Sec.V, we make a few comments on some of the consequences of our findings. In particular, we discuss some relations among the masses of doubly-heavy-light and heavy-light hadrons. Appendix A contains notation and definitions. To make the paper more self-contained, we give a brief review of the $QQq$ system in Appendix B.
\section{Preliminaries}
\renewcommand{\theequation}{2.\arabic{equation}}
\setcounter{equation}{0}
\subsection{A general procedure}

In the presence of light quarks, the B-O potentials can be determined along the lines of lattice QCD. This implies that a mixing analysis based on a correlation matrix is needed \cite{FK}. Its diagonal elements are given by the energies of stationary configurations, whereas the off-diagonal ones describe transitions between those configurations. The potentials are determined by the eigenvalues of this matrix. 

\begin{figure}[htbp]
\centering
\includegraphics[width=9.5cm]{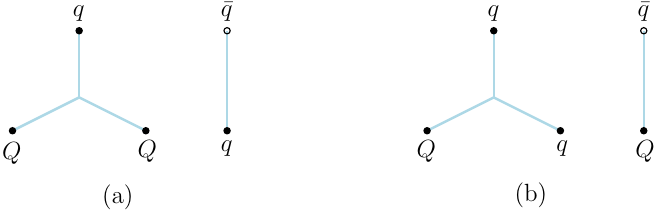}
\caption{{\small Basic string configurations. Three strings may join at a point called the string junction \cite{XA2}. Here and later, non-excited strings are designated by straight lines.}}
\label{c40}
\end{figure}

Now consider string configurations for the $QQqq\bar q$ quark system from the standard viewpoint in four dimensions \cite{XA}. We do so for $N_f=2$, two dynamical flavors of equal mass, but the extension to $N_f=2+1$ is straightforward. First let us look at the simplest disconnected configurations only with the valence quarks. These are the basic configurations shown in Figure \ref{c40}. Each consists of the valence quarks and antiquark connected by the strings and looks like a pair  

 \noindent of non-interacting hadrons. 

\begin{figure}[htbp]
\centering
\includegraphics[width=9.5cm]{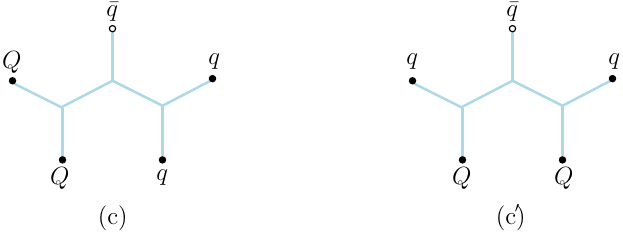}
\caption{{\small Pentaquark configurations.}}
\label{c41}
\end{figure}

\begin{figure*}
\centering
\includegraphics[width=17cm]{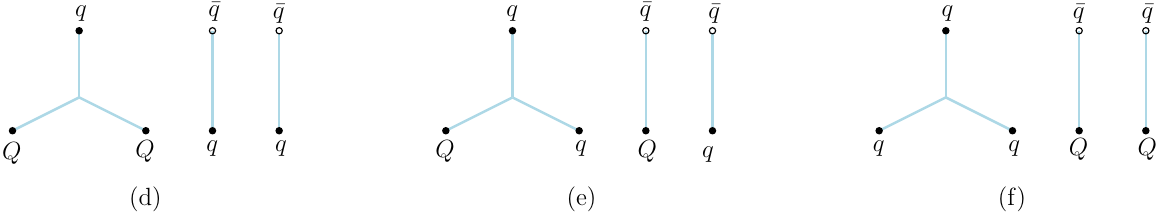}
\caption{{\small String configurations with one virtual pair.}}
\label{c42}
\end{figure*}

To pursue this further, we assume that other configurations are constructed by adding extra string junctions and virtual quark-antiquark pairs to the basic configurations. Intuitively, it is clear that such a procedure will result in configurations of higher energy. And so to some extent, the junctions and pairs can be thought of as kinds of elementary excitations. It turns out that for our purposes we would only need relatively simple configurations. Adding two string junctions to the basic configurations results in the pentaquark configurations of Figure \ref{c41}.\footnote{Since they describe the genuine five-body interactions of quarks, we call them the pentaquark configurations. As we will see in Sec.III, such configurations make the dominant contribution to the potential at small heavy quark separations. Because of this, we will add the word "compact" as a prefix.} On the other hand, adding 

\noindent one $q\bar q$ pair results in the configurations of Figure \ref{c42}. The configurations (d) and (e) are simple 
modifications of the configurations (a) and (b), respectively. The configuration (f) is obtained from those by quark exchange. It is noteworthy that apart from the string junctions and virtual pairs other elementary excitations may be involved. We return to this in Sec.V.

\begin{figure}[htbp]
\centering
\includegraphics[width=13cm]{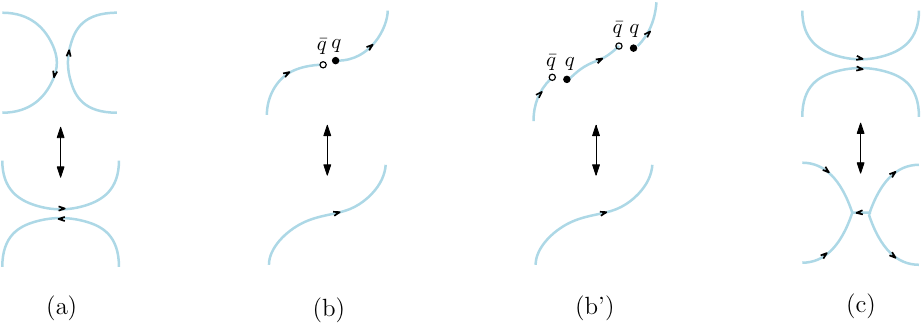}
\caption{{\small Some string interactions: (a) reconnection (rearrangement); (b) breaking; (b') double breaking; (c) junction annihilation.}}
\label{sint}
\end{figure}

The possible transitions between the configurations arise from string interactions. In Figure \ref{sint}, we sketch four
different types of interactions which will be discussed in what follows. This is part of the big picture of QCD strings \cite{XA}. Later we will introduce the notion of a critical separation between the heavy quarks, which characterizes each interaction, that will enable us to shed some light on the physics of QCD strings. 

\subsection{A short account of the five-dimensional string model}

In our study of the $QQqq\bar q$ system, we will use the formalism recently developed in \cite{a-strb}. Although we illustrate it by performing calculations in one of the simplest AdS/QCD models, the formalism is general and, therefore, adaptable to other models as well.

For the purposes of this paper, we consider a five-dimensional Euclidean space with coordinates $t, x^i, r$ and with the metric  

\begin{equation}\label{metric}
ds^2=\ep^{\s r^2}\frac{R^2}{r^2}\Bigl(dt^2+(dx^i)^2+dr^2\Bigr)
\,.
\end{equation}
Such a space represents a deformation of the Euclidean $\text{AdS}_5$ space of radius $R$, with a deformation parameter $\s$. The boundary is at $r=0$ and the so-called soft wall at $r=1/\sqrt{\s}$. Two features make it especially attractive: relative computational simplicity and phenomenological applications. Here let us just mention that the model of \cite{az1} yields a perfect fit to the lattice data obtained for the heavy quark potential \cite{white}.\footnote{For another perfect example, see \cite{a-3qPRD}.} 

As for Feynman diagrams in field theory, we need the building blocks to construct the string configurations of Figures \ref{c40}-\ref{c42} in five dimensions. The first is a Nambu-Goto string governed by the action 

\begin{equation}\label{NG}
S_{\text{\tiny NG}}=\frac{1}{2\pi\alpha'}\int d^2\xi\,\sqrt{\gamma^{(2)}}
\,.
\end{equation}
Here $\gamma$ is an induced metric, $\alpha'$ is a string parameter, and $\xi^i$ are world-sheet coordinates. 

The second is a high-dimensional counterpart of the string junction called the baryon vertex. In the AdS/CFT correspondence this vertex is supposed to be a dynamic object. It is a five brane wrapped on an internal space $\mathbf{X}$ \cite{witten}, and correspondingly the antibaryon vertex is an antibrane. Both objects look point-like in five dimensions. In \cite{a-3qPRD} it was observed that the action for the baryon vertex, written in the static gauge, 

\begin{equation}\label{baryon-v}
S_{\text{vert}}=\tau_v\int dt \,\frac{\ep^{-2\s r^2}}{r}
\,
\end{equation}
yields very satisfactory results, when compared to the lattice calculations of the three-quark potential. Notice that $S_{\text{vert}}$ is given by the worldvolume of the brane if $\tau_v={\cal T}_5R\,\text{vol}(\mathbf{X})$, with ${\cal T}_5$ a brane tension. Unlike AdS/CFT, we treat $\tau_v$ as a free parameter to somehow account for $\alpha'$-corrections as well as possible impact of the other background fields.\footnote{Like in AdS/CFT, one expects an analog of the Ramond-Ramond fields on $\mathbf{X}$.} In the case of zero baryon chemical potential, it is natural to take the action \eqref{baryon-v} also for the antibaryon vertex so that $S_{\bar{\text{vert}}}=S_{\text{vert}}$.

Finally, following \cite{son}, we introduce a background scalar field $\text{T}(r)$ which describes two light quarks of equal mass. In the present context those are at string endpoints in the interior of five-dimensional space. The scalar field couples to the worldsheet boundary as an open string tachyon $S_{\text{q}}=\int d\tau e\,\text{T}$, where $\tau$ is a coordinate on the boundary and $e$ is a boundary metric.\footnote{The use of the term tachyon also seems particularly appropriate in virtue of instability of a QCD string.} In what follows, we consider only a constant field $\text{T}_0$ and worldsheets whose boundaries are straight lines in the $t$ direction. In that case, the action written in the static gauge is simply

\begin{equation}\label{Sq}
S_{\text q}=\text{T}_0R\int dt \frac{\ep^{\oh\s r^2}}{r}
\,.
\end{equation}
One can immediately recognize it as the action of a point particle of mass ${\text T}_0$ at rest.\footnote{For the parameter values used in this paper, the masses of the light quarks can be found by fitting the string breaking distance for the $Q\bar Q$ system to the lattice data of \cite{bulava}. This leads to the result $m_{u/d}=46.6\,\text{MeV}$ \cite{a-stb3q}.}  Clearly, at zero baryon chemical potential the same action also describes the light antiquarks, and hence $S_{\bar{\text q}}=S_{\text q}$. 

It is worth noting a visual analogy between tree level Feynman diagrams and static string configurations. In the language of Feynman diagrams the above building blocks play respectively the roles of propagators, vertices and tadpoles.

\section{The $QQ qq\bar q$-Quark System: string theory analysis in five dimensions}
\renewcommand{\theequation}{3.\arabic{equation}}
\setcounter{equation}{0}

We will now describe the $QQqq\bar q$ system. Our basic approach is as follows. Following the hadro-quarkonium picture \cite{voloshin}, we think of the light quarks/antiquarks as clouds. Therefore, it only makes sense to speak about the average positions of those or, equivalently, the centers of the clouds. The heavy quarks are point-like objects inside the clouds. Our goal is to determine the low-lying B-O potentials as a function of separation between the heavy quarks. 

We start our discussion with the basic string configurations, then continue with the remaining ones, and finally end up with the potentials. A simple intuitive way to see how a configuration looks like in five dimensions is to place it on the boundary of five-dimensional space. A gravitational force pulls the light quarks and strings into the interior, whereas the heavy (static) quarks remain at rest. Mostly this helps, but some exceptions exist. We will see shortly that the shape of several configurations changes with the heavy quark separation that makes the problem more complicated.

\subsection{The disconnected configurations (a) and (b)}

First consider configuration (a). It can be interpreted as a pair of non-interacting hadrons: a doubly heavy baryon and a pion. Clearly, the total energy is just the sum of the rest energies of the baryon and pion if they are infinitely far apart. It is quite surprising that such a factorization also holds at finite separation if one takes the average over the pion (cloud) position \cite{pion-factor}. In what follows, we consider the configurations with pions only in the sense of averaging over all possible pion positions. 

\begin{figure}[H]
\centering
\includegraphics[width=5.25cm]{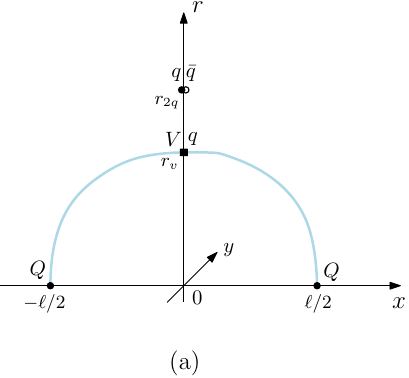}
\hspace{2.5cm}
\includegraphics[width=5.25cm]{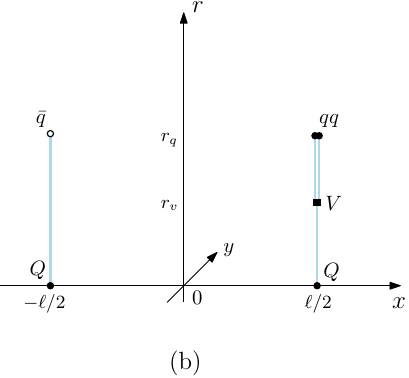}
\caption{{\small Configurations (a) and (b) in five dimensions. The heavy quarks are placed on the boundary at $r=0$ and separated by distance $\ell$. The light quarks, baryon vertices and pion are at $r=\rq$, $r=\rv$ and $r=r_{2q}$, respectively. Generically, the shape of configuration (a) changes with $\ell$. Sketched here is the configuration for intermediate separations (see Fig.\ref{cQQq}).}}
\label{con-ab}
\end{figure}

In five dimensions, the configuration looks like that shown in Figure \ref{con-ab}(a). It consists of two parts. The lower part 
\noindent corresponds to the $QQq$ system and the upper to the pion. The total energy is the sum of the energies of each part 

\begin{equation}\label{Ea}
E^{\text{(a)}}=E_{\QQq}+E_{\qqb}
\,.	
\end{equation}
In the static limit, $E_{\QQq}$ was computed in \cite{a-QQq}. For convenience, we include a brief description of the results in Appendix B. $E_{\qqb}$ was computed in \cite{a-QQbqqb} with the result

\begin{equation}\label{Eqqb}
E_{\qqb}=2\n\sqrt{\g\sigma}
\,,\qquad
\sigma=\ep\g\s
\,.
\end{equation}
Here $\g=\frac{R^2}{2\pi\alpha'}$, $\n=\frac{\text{T}_0 R}{\g}$, and $\sigma$ is the string tension obtained in \cite{az1} from the study of the $Q\bar Q$ system at large pair separation.

Now let us consider configuration (b). Again, the total energy is just the sum of the rest energies. So, 

\begin{equation}\label{Eb}
E^{\text{(b)}}=E_{\Qqb}+E_{\Qqq}
\,.	
\end{equation}
The first term is the rest energy of a heavy-light meson. In \cite{a-strb} it was shown that it can be written as 

\begin{equation}\label{Qqb}
E_{\Qqb}=\g\sqrt{\s}\Bigl({\cal Q}(\qs)+\n \frac{\ep^{\oh \qs}}{\sqrt{\qs}}\Bigr)+c
\,,
	\end{equation}
where the function ${\cal Q}$ is defined in Appendix A and $c$ is a normalization constant. $\qs$ is a solution to the equation 

\begin{equation}\label{q}
\ep^{\frac{q}{2}}+\n(q-1)=0
\,
\end{equation}
in the interval $[0,1]$. Here $q=\s \rq^2$. This equation is nothing else but the force balance equation in the $r$-direction. It is derived by varying the action $S=S_{\text{\tiny NG}}+S_{\text q}$ with respect to $\rq$. 

The second term represents the rest energy of a heavy-light baryon. It was also computed in \cite{a-strb}. In this case, one has

\begin{equation}\label{EQqq}
E_{\Qqq}=\g\sqrt{\s}\Bigl(2{\cal Q}(\qs)-{\cal Q}(\vs)
+2\n \frac{\ep^{\oh \qs}}{\sqrt{\qs}}
+3\k \frac{\ep^{-2\vs}}{\sqrt{\vs}}
\Bigr)+c
\,,
	\end{equation}
with $\vs$ a solution to the equation

\begin{equation}\label{v}
1+3\k(1+4v)\ep^{-3v}=0
\,
\end{equation}
in the interval $[0,1]$. Here $\k=\frac{\tau_v}{3\g}$ and $v=\s\rv^2$. The above equation is the force balance equation in the $r$-direction at $r=\rv$. It is derived by varying the action $S=3S_{\text{\tiny NG}}+2S_{\text q}+S_{\text{vert}}$ with respect to $\rv$. 

We conclude our discussion of the basic configurations with some remarks. First, in \cite{a-QQq} it was shown that in the interval $[0,1]$ Eq.\eqref{v} has solutions if and only if $\k$ is restricted to the range $-\frac{\ep^3}{15} <\k\leq -\frac{1}{4}\ep^{\frac{1}{4}}$. In particular, there exists a single solution $\vs=\frac{1}{12}$ at $\k= -\frac{1}{4}\ep^{\frac{1}{4}}$. Second, the analysis of configuration (b) assumes that $\vs\leq \qs$. Although this is not true for all possible parameter values, it definitely is for those we use to make predictions. Finally, the solutions $\qs$ and $\vs$ describe the light quarks and baryon vertices and, as a consequence, are independent of the heavy quark separation.
\subsection{The connected configuration (c)}

Now let us discuss the pentaquark configuration (c) in five dimensions. In doing so, it is natural to suggest that if a configuration contributes to the ground state, or at least to the first excited state, its shape is dictated by symmetry. For the configuration at hand, there are the two most symmetric cases: 1. All the light quarks are in the middle between the heavy quarks. 2. The light quarks sit on top of each heavy quark and the antiquark is in the middle.\footnote{This is in fact the antiquark-diquark-diquark scheme of \cite{maiani}.} We will see that the former takes place at small enough separations between the heavy quarks, whereas the latter at larger separations, $\ell\gtrsim 0.1\,\text{fm}$.

\subsubsection{Small $\ell$}

\begin{figure}[H]
\centering
\includegraphics[width=6.6cm]{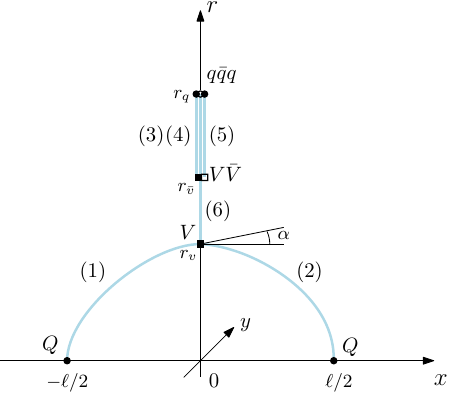}
\caption{{\small The pentaquark configuration (c) for very small $\ell$. The light quarks and baryon vertices are placed on the $r$-axis. Here and later, $\alpha$ indicates the tangent angle at the endpoint of string (1).}}
\label{c1}
\end{figure}

For small separations, the corresponding string configuration is depicted in Figure \ref{c1}.\footnote{We are jumping ahead slightly, as we have not discussed the configuration with $V\bar V$ spatially splitted apart. In that case, one has to extremize the total action with respect to both vertex positions. A technical analysis goes along the same lines as those for $Qqq$ in \cite{a-strb} and leads to the configuration of Figure \ref{c1}.} It is obtained by placing the 
\noindent standard configuration of Figure \ref{c41}(c) on the boundary of five-dimensional space. The light quarks are indeed in the middle between the heavy ones. Here $\rq$, $\rv$, and $\rvb$ are assumed to satisfy the following condition: $\rq>\rvb >\rv$.\footnote{As we will see shortly, it holds for the parameter values we use in this work. }  

The total action is the sum of the Nambu-Goto actions plus the actions for the vertices and light quarks

\begin{equation}\label{action-cs}
S=\sum_{i=1}^6 S_{\text{\tiny NG}}^{(i)}+3S_{\text{vert}}+3S_{\text q}
\,.
\end{equation}
We pick the static gauge $\xi^1=t$ and $\xi^2=r$ for the Nambu-Goto actions and consider the $x^{(i)}$'s as a function of $r$. Then the boundary conditions take the form

\begin{equation}\label{boundary-s}
x^{(1;2)}(0)=\mp\oh\ell\,,
\qquad
x^{(1,2,6)}(\rv)=x^{(3,4,5,6)}(\rvb)=x^{(3,4,5)}(\rq)=0\,,
\end{equation}
and the action becomes\footnote{We drop the subscript $(i)$ when it does not cause confusion.}

\begin{equation}\label{action-s2}
S=\g T
\biggl(
2\int_{0}^{\rv} \frac{dr}{r^2}\,\ep^{\s r^2}\sqrt{1+(\partial_r x)^2}\,\,
+
\int_{\rv}^{\rvb} \frac{dr}{r^2}\,\ep^{\s r^2}
+
3\int_{\rvb}^{\rq} \frac{dr}{r^2}\,\ep^{\s r^2}
+
3\k\,\frac{\ep^{-2\s\rv^2}}{\rv}
+
6\k\,\frac{\ep^{-2\s\rvb^2}}{\rvb}
+
3\n\frac{\ep^{\frac{1}{2}\s\rq^2}}{\rq}
\,\biggr)
\,,
\end{equation}
where $\partial_rx=\frac{\partial x}{\partial r}$ and $x^{(3,4,5,6)}=const$. The integrals represent the contributions of the strings, and the remaining terms the contributions of the vertices and light quarks. 

To find a stable configuration, one has to extremize the action with respect to $x(r)$ describing the profiles of strings (1) and (2), and, in addition, with respect to $\rv$, $\rvb$, and $\rq$ describing the locations of the vertices and light quarks. As explained in Appendix B of \cite{a-stb3q}, varying with respect to $x^{(1)}$ results in the following expression for the separation distance and the energy of the configuration:

\begin{equation}\label{l-s}
\ell=\frac{2}{\sqrt{\s}}{\cal L}^+(\alpha,v)
\,,
\end{equation}
 \begin{equation}\label{E-s}
E^{\text{(c)}}=\frac{S}{T}=E_{\Q2q3}=\g\sqrt{\s}
\biggl(
2{\cal E}^+(\alpha,v)
+
3{\cal Q}(q)-2{\cal Q}(\bar v)-{\cal Q}(v)
+
3\k\frac{\ep^{-2v}}{\sqrt{v}}
+
6\k\frac{\ep^{-2\bar v}}{\sqrt{\bar v}}
+
3\n\frac{\ep^{\oh q}}{\sqrt{q}}
\biggr)
+2c
\,.
\end{equation}
We have also used the fact that $\int_a^b\frac{dx}{x^2}\ep^{cx^2}=\sqrt{c}\bigl({\cal Q}(cb^2)-{\cal Q}(ca^2)\bigr)$. Here $v=\s\rv^2$, $\bar v=\s\rvb^2$, and the functions ${\cal L}^+$ and ${\cal E}^+$ are defined in Appendix A. 

It is easy to see that varying the action with respect to $\rq$ and $\rvb$ leads to Eq.\eqref{q} and Eq.\eqref{v} (with $v$ replaced by $\bar v$). On the other hand, varying with respect to $\rv$ leads to 

\begin{equation}\label{alpha1}
\sin\alpha=\oh\bigl(1+3\k(1+4v)\ep^{-3v}\bigr)
\,.	
\end{equation}
All these equations are nothing else but the force balance equations in the $r$-direction.

Combining Eqs.\eqref{E-s}-\eqref{alpha1} with Eq.\eqref{EQQqs}, we can write the energy as

\begin{equation}\label{Emaiani}
E_{\Q2q3}=E_{\QQq}+
2\g\sqrt{\s}
\biggl(
{\cal Q}(\qs)-{\cal Q}(\vs)
+
3\k\frac{\ep^{-2\vs}}{\sqrt{\vs}}
+
\n\frac{\ep^{\oh \qs}}{\sqrt{\qs}}
\biggr)
\,.
\end{equation}
Here $E_{\QQq}$ is the energy of the configuration shown in Figure \ref{cQQq}(s). $\qs$ and $\vs$ are solutions respectively to Eqs.\eqref{q} and \eqref{v} in the interval $[0,1]$. In four dimensions, this formula assumes a structure like $QQ(qq\bar q)$ with a color-triplet $(qq\bar q)$.

Thus, the energy of the configuration is given in parametric form by $E_{\Q2q3}=E_{\Q2q3}(v)$ and $\ell=\ell(v)$. The parameter takes values in the interval $[0,\vs]$.

\subsubsection{Larger $\ell$}

\begin{figure}[htbp]
\centering
\includegraphics[width=6.55cm]{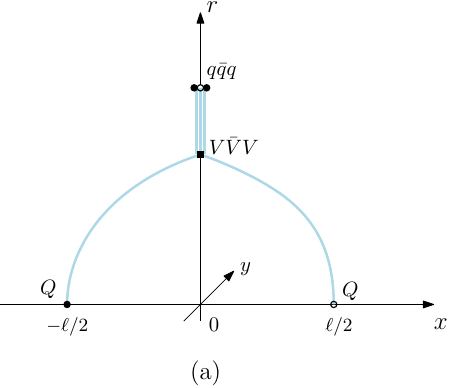}
\hspace{2.75cm}
\includegraphics[width=6.5cm]{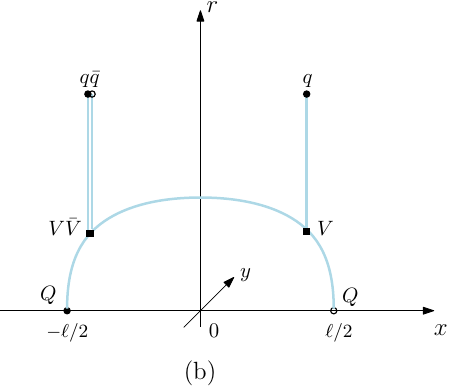}
\caption{{\small Some other pentaquark configurations.}}
\label{c2unstable}
\end{figure}

\begin{figure}[htbp]
\centering
\includegraphics[width=7.45cm]{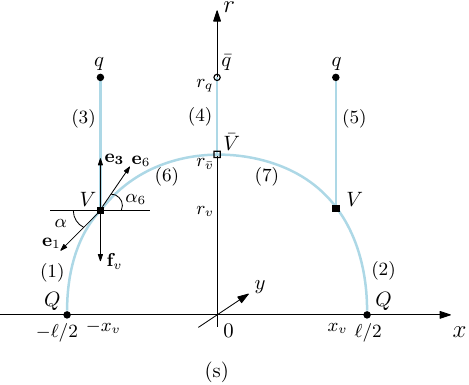}
\caption{{\small The pentaquark configuration (c) for small $\ell$ (but larger than $\ell(\vs))$. It is labeled by (s), like the corresponding configuration for the $QQq$-system (see Fig. \ref{cQQq}). The forces exerted on the baryon vertex are depicted by the arrows. $\alpha_6$ is the tangent angle at the endpoint of string (6).}}
\label{c2s}
\end{figure}

A simple numerical analysis shows that $\ell$ is an increasing function of $v$ vanishing at the origin.\footnote{This is indeed the case for the parameter values we use.}  Therefore, at some separation between the heavy quarks the baryon vertex reaches the point $r=\rvb$ where it meets the other vertices, as shown in Figure \ref{c2unstable}(a).\footnote{For $\ell<\ell(\vs)$, this configuration has higher energy than the configuration of Figure \ref{c1}. Its four-dimensional counterpart is sketched in Figure \ref{c41}(c').} This  defines the upper bound on the range of $v$ in \eqref{l-s} and \eqref{E-s}. We can continue with a $V\bar VV$ object until we approach infinite separation.  
Why do we need to be concerned with such a configuration? The reason is that from the string theory perspective it is natural to expect that the brane-antibrane annihilation or, in other words, the string junction annihilation occurs if the positions of the vertices coincide or close enough to each other. This could lead to instability. However, there is a natural way out of it: employing another configuration in which the vertices are spatially separated, as depicted in Figure \ref{c2s}.\footnote{Another reason is that it is most symmetric. For comparison, see the configuration shown in Figure \ref{c2unstable}(b), which does not have reflection symmetry.}

This configuration is governed by the action 

\begin{equation}\label{action-cl}
S=\sum_{i=1}^7 S_{\text{\tiny NG}}^{(i)}+3S_{\text{vert}}+3S_{\text q}
\,.
\end{equation}
In the static gauge the boundary conditions are 

\begin{equation}\label{bond-cl}
x^{(1;2)}(0)=\mp\frac{\ell}{2}\,,
\qquad
x^{(1,3,6;2,5,7)}(\rv)=x(\rq)^{(3;5)}=\mp x_v\,,
\qquad
x^{(4,6,7)}(\rv)=x^{(4)}(\rq)=0
\,,
\end{equation}
and the action reads 

\begin{equation}\label{action-cl2}
\begin{split}
S=&\g T
\biggl(
\int_{0}^{\rv} \frac{dr}{r^2}\,\ep^{\s r^2}\sqrt{1+(\partial_r x)^2}
+\int_{\rv}^{\rq} \frac{dr}{r^2}\,\ep^{\s r^2}
+\int_{\rv}^{\rvb} \frac{dr}{r^2}\,\ep^{\s r^2}\sqrt{1+(\partial_r x)^2}
+ 
3\k\,\frac{\ep^{-2\s\rv^2}}{\rv}
+\n\frac{\ep^{\frac{1}{2}\s\rq^2}}{\rq}
\,\biggr)
+(x\rightarrow -x)\\
+
&\g T
\biggl(
\int_{\rvb}^{\rq} \frac{dr}{r^2}\,\ep^{\s r^2}
+
3\k\,\frac{\ep^{-2\s\rvb^2}}{\rvb}
+\n\frac{\ep^{\frac{1}{2}\s\rq^2}}{\rq}
\,
\biggr)
\,.
\end{split}
\end{equation}
Here we set $x^{(3,4)}=const$. Because of the reflection symmetry of the configuration, only the $x<0$ part of the action is written explicitly in the first line of \eqref{action-cl2}. The integrals correspond respectively to the contributions of strings (1), (3), and (6). The second line represents the $x=0$ part of the action. 

Given the action \eqref{action-cl2}, it is straightforward to extremize it   with respect to the $x(r)$'s describing the profiles of strings and with respect to $x_v$, $\rv$, $\rvb$, and $\rq$ describing the locations of the baryon vertices and light quarks. We begin with $x_v$ and $\rv$. In this case, the result can be conveniently written in a vector form as 

\begin{equation}\label{V-fbe}
\mathbf{e}_1+\mathbf{e}_3+\mathbf{e}_6+\mathbf{f}_v=0
\,,
\end{equation}
where $\mathbf{e}_1=\g w(\rv)(-\cos\alpha,-\sin\alpha)$, $\mathbf{e}_3=\g w(\rv)(0,1)$, $\mathbf{e}_6=\g w(\rv)(\cos\alpha_6,\sin\alpha_6)$, and $\mathbf{f}_v=(0,-3\g\k\,\partial_{\rv}\frac{\ep^{-2\s\rv^2}}{\rv})$, with  $\alpha_i\leq\frac{\pi}{2}$. This is the force balance equation at the position of the baryon vertex (see Fig. \ref{c2s}). Its $x$-component reduces simply to  

\begin{equation}\label{Vx-fbe}
\cos\alpha-\cos\alpha_6 =0
\,. 
\end{equation}
The equation has a trivial solution $\alpha_6=\alpha$. "Trivial" means that in this case the strings (1) and (6) are smoothly glued together to form a single string. The vertex in turn does not affect the string.\footnote{In fact, this is true only for $v=\vs$.} If so, then the $r$-component becomes equivalent to Eq.\eqref{v}. As a result, a significant simplification of the expression \eqref{action-cl2} is achieved. Now it takes the form  

\begin{equation}\label{action-cl3}
S=\g T
\biggl(
2\int_{0}^{\rvb} \frac{dr}{r^2}\,\ep^{\s r^2}\sqrt{1+(\partial_r x)^2}
+
\int_{\rvb}^{\rq} \frac{dr}{r^2}\,\ep^{\s r^2}
+
2\int_{\rv}^{\rq} \frac{dr}{r^2}\,\ep^{\s r^2}
+ 
6\k\,\frac{\ep^{-2\s\rv^2}}{\rv}
+
3\k\,\frac{\ep^{-2\s\rvb^2}}{\rvb}
+
3\n\frac{\ep^{\frac{1}{2}\s\rq^2}}{\rq}
\,
\biggr)
\,.
\end{equation}

\begin{figure}[htbp]
\centering
\includegraphics[width=6.9cm]{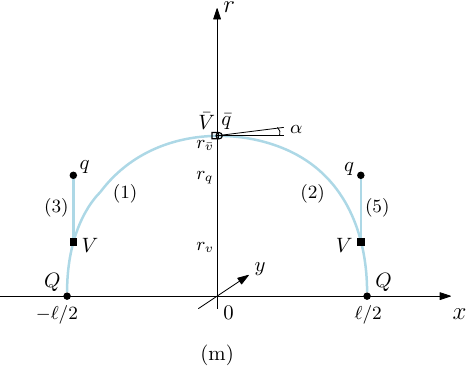}
\hspace{2.5cm}
\includegraphics[width=6.9cm]{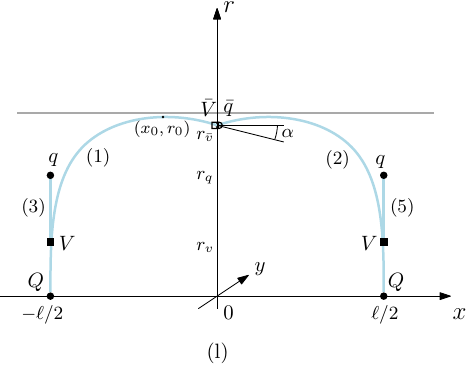}
\caption{{\small Static configurations for intermediate (m) and large (l) separations. The gray horizontal line in (l) represents the soft wall at $r=1/\sqrt{\s}$. For string (1) a turning point is located at $(x_0,r_0)$.}}
\label{c2ml}
\end{figure}

Varying the action with respect to $\rq$ and $\rvb$ results in Eqs.\eqref{q} and \eqref{alpha1}, with $v$ replaced by $\bar v$. The same reasoning that led to \eqref{l-s} gives

\begin{equation}\label{l-s2}
\ell=\frac{2}{\sqrt{\s}}{\cal L}^+(\alpha,\bar v)
\,.
\end{equation}
Similarly the obvious analog of Eq.\eqref{E-s} holds for the energy

\begin{equation}\label{E-s2}
E_{\Q2q3}=\g\sqrt{\s}
\biggl(
2{\cal E}^+(\alpha,\bar v)
+
3{\cal Q}(\qs)
-
2{\cal Q}(\vs)
-
{\cal Q}(\bar v)
+
6\k\frac{\ep^{-2\vs}}{\sqrt{\vs}}
+
3\k\frac{\ep^{-2\bar v}}{\sqrt{\bar v}}
+
3\n\frac{\ep^{\oh\qs}}{\sqrt{\qs}}
\biggr)
+2c
\,.
\end{equation}
Given this, we learn by using \eqref{EQQqs} that the same formula as in \eqref{Emaiani} is also valid in the interval $[\vs,\qs]$. An important point is that now it assumes an antiquark-diquark-diquark $\bar q[Qq][Qq]$ structure. The reason for the upper bound is that the antibaryon vertex goes up as the separation between the heavy quarks is increased. This continues until it reaches the light quark whose position is independent of the separation. So, the string configuration becomes that shown in Figure \ref{c2ml}(m). 

Again, a short numerical calculation shows that $\ell$ reaches its maximum value at $v=\qs$ but remains finite. Because of this, for larger separations one needs to analyze the configuration of Figure \ref{c2ml}(m). By essentially the same arguments that we have given for the configuration of Figure \ref{c2s}, the strings (1) and (2), which are stretched between the heavy quarks and antibaryon vertex, are smooth. In that case, the action can be obtained from \eqref{action-cl3} by simply dropping the contribution of string (4) out. So,  

\begin{equation}\label{action-cm}
S=\g T
\biggl(
2\int_{0}^{\rvb} \frac{dr}{r^2}\,\ep^{\s r^2}\sqrt{1+(\partial_r x)^2}
+
2\int_{\rv}^{\rq} \frac{dr}{r^2}\,\ep^{\s r^2}
+ 
6\k\,\frac{\ep^{-2\s\rv^2}}{\rv}
+
3\k\,\frac{\ep^{-2\s\rvb^2}}{\rvb}
+
2\n\frac{\ep^{\frac{1}{2}\s\rq^2}}{\rq}
+
\n\frac{\ep^{\frac{1}{2}\s\rvb^2}}{\rvb}
\,
\biggr)
\,.
\end{equation}
As before, varying the action with respect to $\rq$ leads to Eq.\eqref{q} and with respect to $\rv$ to Eq.\eqref{v}. But when one varies it with $\rvb$, the resulting formula can be written as  

\begin{equation}\label{alpham}
\sin\alpha=\oh\bigl(
\n (1-\bar v)\ep^{-\oh \bar v}+3\k(1+4\bar v)\ep^{-3\bar v}
\bigr)
\,.
\end{equation}
Since $\alpha$ is non-negative by construction, this places an upper bound on $\bar v$. It is defined by the equation $\alpha=0$, or equivalently, by

\begin{equation}\label{v0}
\n (1-\bar v)+3\k(1+4\bar v)\ep^{-\frac{5}{2}\bar v}=0
\,.
\end{equation}
We denote the solution of this equation in the interval $[0,1]$ as $\Vz$.

Clearly, Eq.\eqref{l-s2} holds for the heavy quark separation. It follows then from \eqref{action-cm} that the energy of the configuration is  

\begin{equation}\label{E-m}
E_{\Q2q3}=\g\sqrt{\s}
\biggl(
2{\cal E}^+(\alpha,\bar v)
+
2{\cal Q}(\qs)
-
2{\cal Q}(\vs)
+
6\k\frac{\ep^{-2\vs}}{\sqrt{\vs}}
+
3\k\frac{\ep^{-2\bar v}}{\sqrt{\bar v}}
+
2\n\frac{\ep^{\oh \qs}}{\sqrt{\qs}}
+
\n\frac{\ep^{\oh \bar v}}{\sqrt{\bar v}}
\biggr)
+2c
\,.
\end{equation}
It is straightforward now, using the formula \eqref{EQQqm}, to see that the relation between $E_{\Q2q3}$ and $E_{\QQq}$ is valid in the interval $[\qs,\Vz]$. 

Since $\ell(\Vz)$ is finite, the question arises of whether larger separations can be reached. The short answer to this question is yes, they can. For this, one needs to  consider a change of the sign of $\alpha$. If so, then the configuration profile becomes convex near $x=0$, as depicted in Figure \ref{c2ml}(l). For some $\alpha$ the strings reach the soft wall that corresponds to infinite separation between the heavy quarks.  

The configuration is governed by the same action as before. The expressions for the separation distance and energy are simply obtained by respectively replacing ${\cal L}^+$ and ${\cal E}^+$ with ${\cal L}^-$ and ${\cal E}^-$, as follows from the analysis in Appendix B of \cite{a-stb3q}. So, we have

\begin{equation}\label{l-cl}
\ell=
\frac{2}{\sqrt{\s}}{\cal L}^-(\lambda,\bar v)
\,
\end{equation}
and 
\begin{equation}\label{E-cl}
E_{\Q2q3}=\g\sqrt{\s}
\biggl(
2{\cal E}^-(\lambda,\bar v)
+
2{\cal Q}(\qs)
-
2{\cal Q}(\vs)
+
6\k\frac{\ep^{-2\vs}}{\sqrt{\vs}}
+
3\k\frac{\ep^{-2\bar v}}{\sqrt{\bar v}}
+
2\n\frac{\ep^{\oh \qs}}{\sqrt{\qs}}
+
\n\frac{\ep^{\oh \bar v}}{\sqrt{\bar v}}
\biggr)
+2c
\,.
\end{equation}
The functions ${\cal L}^-$ and ${\cal E}^-$ are given in Appendix A. The dimensionless variable $\lambda$ is defined by $\lambda=\s r_0^2$, with $r_0=\max r(x)$ as shown in Figure \ref{c2ml}(l). Importantly, $\lambda$ can be expressed in terms of $\bar v$ as \cite{a-QQq}  

\begin{equation}\label{lambda}
\lambda=-\text{ProductLog}
\biggl[-\bar v\ep^{-\bar v}
\Bigl(1-\frac{1}{4}\Bigl(3\k(1+4\bar v)\ep^{-3\bar v}
+
\n(1-\bar v)\ep^{-\oh \bar v}\Bigr)^2
\,\Bigr)^{-\frac{1}{2}}
\,\biggr]
\,.
\end{equation}
Here $\text{ProductLog}(z)$ is the principal solution for $w$ in $z=w\,\ep^w$ \cite{wolfram}.

The parameter $\bar v$ now varies from $\Vz$ to $\Vo$. The upper bound is found by solving the equation $\lambda=1$, or equivalently the equation

\begin{equation}\label{v1-QQq}
2\sqrt{1-\bar v^2\ep^{2(1-\bar v)}}+3\k(1+4\bar v)\ep^{-3\bar v}+\n (1-\bar v)\ep^{-\oh \bar v}=0
\,,
\end{equation}
in the interval $[0,1]$. The reasoning for this is that since ${\cal L}^-(\lambda,\bar v)$ is singular at $\lambda=1$, $\Vo$ corresponds to infinite separation between the heavy quarks.

We conclude the discussion of configuration (c) with a brief summary of our analysis. First, $E_{\Q2q3}$ is a piecewise function of $\ell$. The different parts of its domain are described by the different string configurations. Second, for $\ell>\ell(\vs)$ our model provides an explicit realization of the antiquark-diquark-diquark scheme of the pentaquark \cite{maiani}. In that case the formula \eqref{Emaiani} establishes the relation between the energies of a compact pentaquark and a doubly heavy baryon in the heavy quark limit. 
\subsubsection{The limiting cases}

It is instructive to examine an asymptotic behavior of $E_{\Q2q3}$ for small and large $\ell$. These are easy to understand because the relation \eqref{Emaiani} between $E_{\Q2q3}$ and $E_{\QQq}$ is valid for all separations.

The behavior for small $\ell$ can be read off from Eq.\eqref{EQQq-small}. We get exactly what we expect from heavy quark-diquark symmetry \cite{wise}, namely

\begin{equation}\label{penta-factor}
E_{\Q2q3}(\ell)=E_{\QQ}(\ell)+E_{\Qqbqq}
\,.
\end{equation}	
Here $E_{\QQ}$ is the heavy quark-quark potential (in the antitriplet channel) and $E_{\Qqbqq}$ is the rest energy of a heavy-light tetraquark. At zero baryon chemical potential $E_{\Qqbqq}=E_{\Qqqq}$, where  

\begin{equation}\label{EQqqq}
E_{\Qqqq}=3\g\sqrt{\s}
\biggl({\cal Q}(\qs)-\frac{2}{3}{\cal Q}(\vs)
+2\k\frac{\ep^{-2{\vs}}}{\sqrt{\vs}}
+\n\frac{\ep^{\oh\qs}}{\sqrt{\qs}}\,
\biggr)
+c
\,,
\end{equation}
as described in \cite{a-QQqq}. 

Similarly, the behavior for large $\ell$ can be read off from Eq.\eqref{EQQq-large}. So, we have 

\begin{equation}\label{penta-large}
	E_{\Q2q3}=\sigma\ell
	+
	2\g\sqrt{\s}
\biggl(
{\cal Q}(\qs)-{\cal Q}(\vs)
+
3\k\frac{\ep^{-2\vs}}{\sqrt{\vs}}
+
\n\frac{\ep^{\oh \qs}}{\sqrt{\qs}}
-
I_{\QQq}
\biggr)
+2c+o(1)
	\,.
\end{equation}
Here $\sigma$ is the same string tension as in \eqref{Eqqb}. 

\subsubsection{Gluing the pieces together}

Now let us discuss the gluing of all the branches of $E_{\Q2q3}(\ell)$. For this, we need to specify the model parameters. Here we use one of the two parameter sets suggested in \cite{a-strb} which is mainly a result of fitting the lattice QCD data to the string model we are considering. The value of $\s$ is fixed from the slope of the Regge trajectory of $\rho(n)$ mesons in the soft wall model with the geometry \eqref{metric}, and as a result, one gets $\s=0.45\,\text{GeV}^2$ \cite{a-q2}. Then, fitting the value of the string tension $\sigma$ defined in \eqref{Eqqb} to its value in \cite{bulava} gives $\g=0.176$. The parameter $\n$ is adjusted to reproduce the lattice result for the string breaking distance in the $Q\bar Q$ system. With $\boldsymbol{\ell}_{\QQb}=1.22\,\text{fm}$ for the $u$ and $d$ quarks \cite{bulava}, one gets $\n=3.057$ \cite{a-strb}. 

In principle, the value of $\k$ could be adjusted to fit the lattice data for the three-quark potential, as done in \cite{a-3qPRD} for pure $SU(3)$ gauge theory, but there are no lattice data available for QCD with two light quarks. There are still two special options: $\k=-0.102$ motivated by phenomenology\footnote{Note that $\k=-0.102$ is a solution to the equation $\alpha_{\QQ}(\k)=\oh\alpha_{\QQb}$ which follows from the phenomenological rule $E_{\QQ}(\ell)=\oh E_{\QQb}(\ell)$ in the limit $\ell\rightarrow 0$.} and $\k=-0.087$ obtained from the lattice data for pure gauge theory \cite{a-3qPRD}. However, both are out of the range of allowed values for $\k$ as follows from the analysis of Eq.\eqref{v}. In this situation it is reasonable to pick $\k=\km$, as the closest to those. 

Having fixed the model parameters, we can immediately perform some simple but important calculations. First let us check that $\qs>\vs$. That is, our construction of the string configurations makes sense. From Eqs.\eqref{q} and \eqref{v}, we find that $\qs=0.566$ and $\vs=\frac{1}{12}$, as desired. In addition, from \eqref{v0} and \eqref{v1-QQq}, we get $\Vz=0.829$ and $\Vo=0.930$. Second, with this value of $v$, we can immediately calculate the smallest separation between the heavy quarks for the configuration shown in Figure \ref{c2s}. So, $\ell(\vs)=0.106\,\text{fm}$. It is quite surprising that the antiquark-diquark-diquark scheme of \cite{maiani} arises at such small separations.

\begin{figure}[htbp]
\centering
\includegraphics[width=8.25cm]{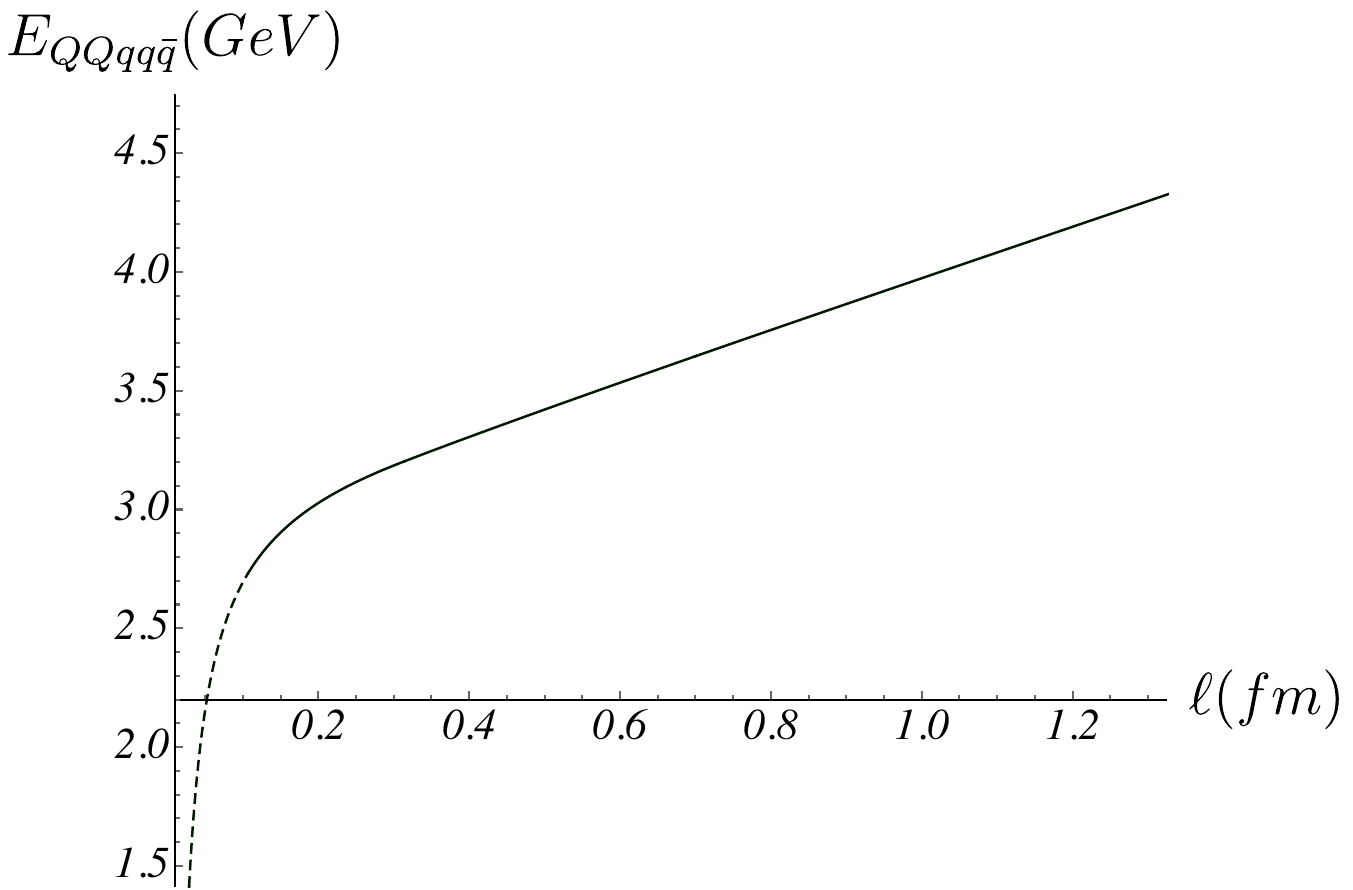}
\caption{{\small $E_{\Q2q3}$ vs $\ell$. The dashed curve corresponds to the configuration of Figure \ref{c1}, whereas the solid curve to the configurations of Figures \ref{c2s} and \ref{c2ml} for which the the antiquark-diquark-diquark scheme holds. Here and later, $c=0.623\,\text{GeV}$.}}
\label{Plotc}
\end{figure}

It is now straightforward to plot $E_{\Q2q3}$ as a function of $\ell$. The result is shown in Figure \ref{Plotc}. From this Figure it is seen that all the pieces of the function  
are smoothly glued together. Moreover, $E_{\Q2q3}(\ell)$ is with good approximation linear for separations larger than $0.4\,\text{fm}$. 

\subsection{The disconnected configurations (d)-(f)}

\begin{figure}[htbp]
\centering
\includegraphics[width=5.1cm]{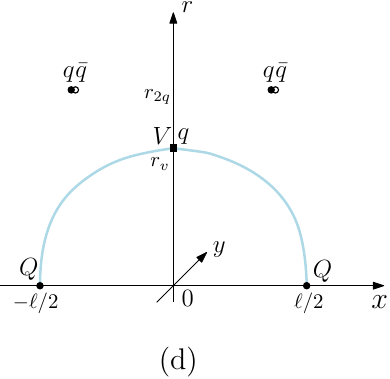}
\hspace{1.12cm}
\includegraphics[width=5.1cm]{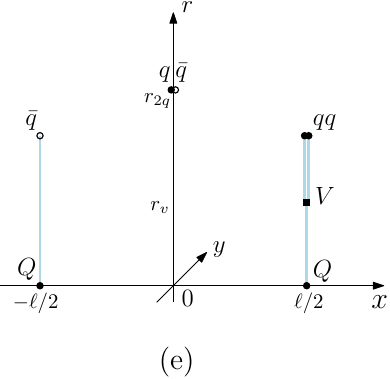}
\hspace{1.12cm}
\includegraphics[width=5.1cm]{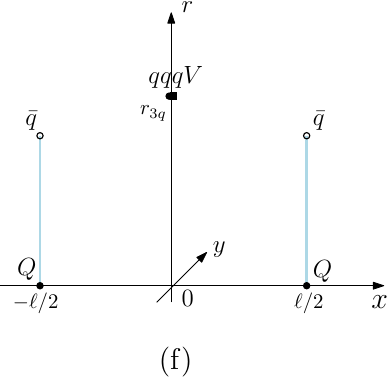}
\caption{{\small Configurations (d), (e), and (f) in five dimensions. The heavy quarks 
in (d) are placed at an intermediate separation distance (see Fig.\ref{cQQq}).}}
\label{con-def}
\end{figure}

We start with configuration (d). It is obtained from configuration (a) by adding a $q\bar q$ pair (pion). So, by analogy with Figure \ref{con-ab}(a), we place the pion at $r=r_{2q}$. The resulting configuration is shown schematically in Figure \ref{con-def}(d). 
Although there are no calculations on the lattice for two pions, we will assume that adding another pion and averaging over its position, leads to an energy increase by $E_{\qqb}$. If so, then the energy is 

\begin{equation}\label{Ed}
E^{\text{(d)}}=E_{\QQq}+2E_{\qqb}
\,,
\end{equation}
with $E_{\QQq}$ described in Appendix B and $E_{\qqb}$ given by \eqref{Eqqb}.

In a similar way, configuration (e) is obtained from configuration (b). It is as shown in Figure \ref{con-def} (e). By the same assumption that we have made previously in our treatment of configuration (d), the energy is given by 

\begin{equation}\label{Ee}
E^{\text{(e)}}=E_{\Qqb}+E_{\Qqq}+E_{\qqb}
\,,
\end{equation}
where the rest energies were defined in Eqs. \eqref{Eqqb}, \eqref{Qqb}, and \eqref{EQqq}.

From the five-dimensional perspective, configuration (f) looks like that shown in Figure \ref{con-def}(f). It consists of the two heavy-light mesons which we have discussed above and a three-quark state (nucleon). Such a state requires some explanation. In the static limit it is natural to guess, by analogy with the case of $q\bar q$, that all the strings connecting the quarks collapse into a point as shown in the Figure.\footnote{This still leaves another choice in which the strings don't collapse, but get stretched along the $r$-axis so that the quarks and vertex are spatially separated. The problem with such a configuration is that it is unstable at $\k=-\frac{1}{4}\ep^{\frac{1}{4}}$.} If so, then the corresponding action is just the sum of the actions of the vertex and light quarks. Explicitly, 

\begin{equation}\label{action-nucl}
S=3\frac{\g T}{\rqqq}
\Bigl(
\k \ep^{-2\s\rqqq^2}
+
\n\ep^{\frac{1}{2}\s\rqqq^2}
\Bigr)
\,.
\end{equation}
The force balance equation obtained by varying $\rqqq$ in \eqref{action-nucl} is 

\begin{equation}\label{3q}
\k(1+4q_3)+\n(1-q_3)\ep^{\frac{5}{2}q_3}=0
\,.
\end{equation}	
Here $q_3=\s\rqqq^2$. The nucleon rest energy is thus

\begin{equation}\label{nucl2}
	E_{\nucl}=
	3\g\sqrt{\frac{\s}{\qsn}}
	\Bigl(
\k\ep^{-2\qsn}
+\n\ep^{\oh\qsn}
\Bigr)
	\,, 
\end{equation}
where $\qsn$ is a solution to Eq.\eqref{3q} in the interval $[0,1]$.\footnote{ Numerically, $\qsn=0.953$ at $\k=\km$ and $\n=3.057$.}

This configuration can be interpreted as a pair of mesons in a nucleon cloud. Like in the case of configuration (e), we assume averaging over a nucleon position and therefore expect that 

\begin{equation}\label{Ef}
E^{\text{(f)}}=2E_{\Qqb}+E_{\nucl}
\,,
\end{equation}
with $E_{\Qqb}$ given by Eq.\eqref{Qqb}.

\subsection{What we have learned}

\begin{figure}[htbp]
\centering
\includegraphics[width=9cm]{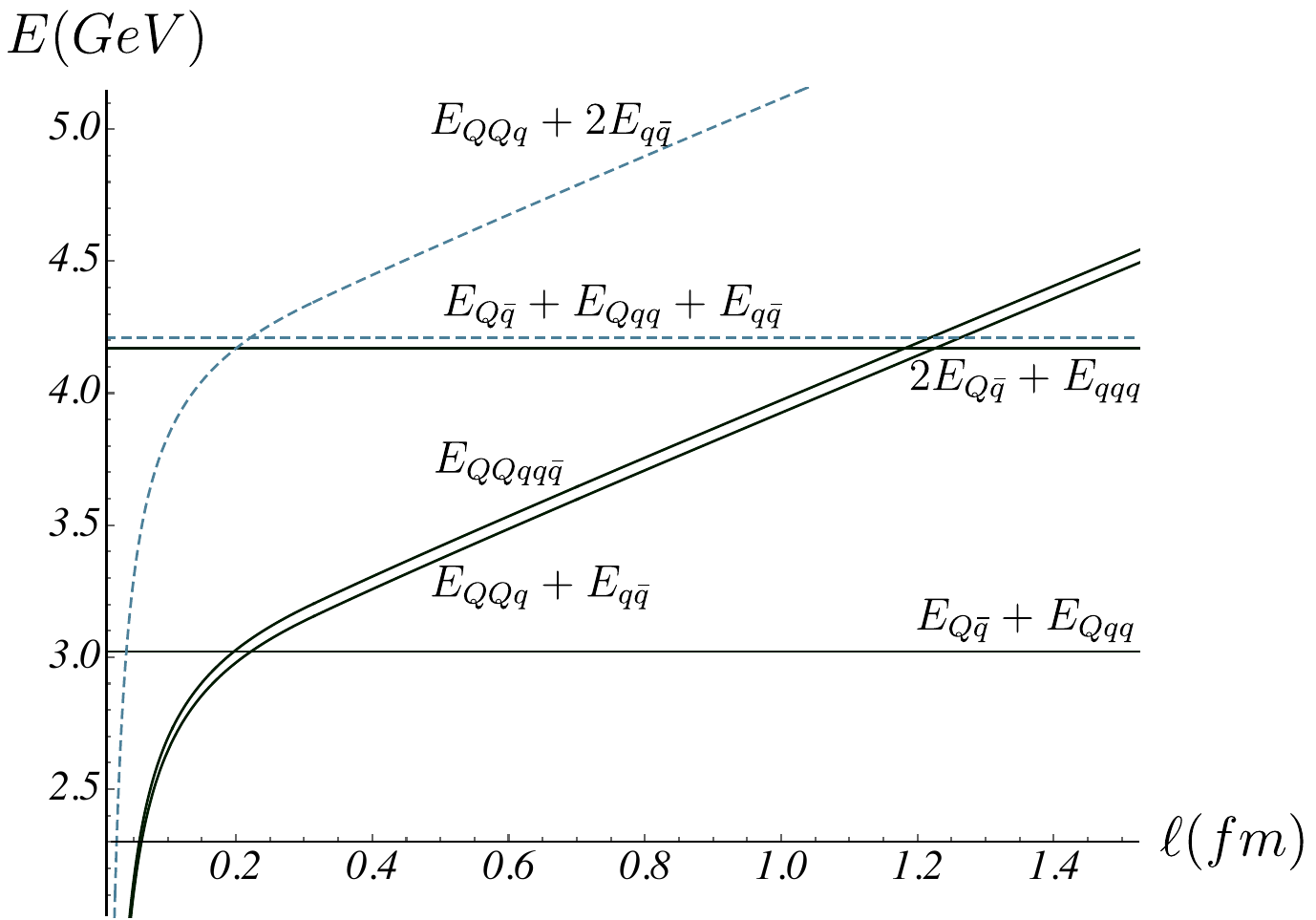}
\caption{{\small Various $E$ vs $\ell$ plots. The curves relevant for $V_0$ and $V_1$ are depicted in solid black.}}
\label{all-L}
\end{figure}

It is instructive to see concretely how the energies of the configurations depend on the separation between the heavy quarks and which configurations contribute to the two low-lying B-O potentials. In Figure \ref{all-L} we plot the $E$'s for our 

\noindent parameter values. These show that the energies of the ground and first excited states are determined only by the contributions from configurations (a)-(c) and (f). In other words, $V_0=\min\{E_{\QQq}+E_{\qqb}, E_{\Qqb}+E_{Qqq}\}$ and $V_1=\min\{E_{\Q2q3},E_{\Qqb}+E_{\Qqq}, E_{\QQq}+E_{\qqb},2E_{\Qqb}+E_{\nucl}\}$.\footnote{Note that in these formulas, one has first to take a minimum to get $V_0$ and then, with that in mind, $V_1$.} 

The essential feature of the pattern of Figure \ref{all-L} is the emergence of three length scales which separate different configurations, or in other words different descriptions. This enables us to better understand how the quarks are organized inside the $QQqq\bar q$ system. 

The first is a scale which refers to the process of string reconnection. It goes like this: $QQq+q\bar q\rightarrow Qqq+Q\bar q$ for $V_0$ and vice versa for $V_1$, if $\ell$ is increased. In the case of $V_0$ the physical meaning of this scale is that the system can be though of as a doubly heavy baryon in a pion cloud for smaller quark separations and, respectively, as a pair of heavy-light hadrons for larger ones. To make this more quantitative, we define a critical separation distance $l_{\Qq}$ by 

\begin{equation}\label{lQqb}
	E_{\QQq}(l_{\Qq})+E_{q\bar q}=E_{\Qqq}+E_{\Qqb}
	\,.
	\end{equation}
If reconnection occurs at small $\ell$ (as in Fig.\ref{all-L}), then by using \eqref{EQQq-small} together with \eqref{Eqqb}, \eqref{Qqb} and \eqref{EQqq}, we get
\begin{equation}\label{lQq-small}
l_{\Qq}=
	\frac{\g\sqrt{\s}}{\boldsymbol{\sigma}_{\QQ}}
	\Bigl({\cal Q}(\qs)-\oh{\cal Q}(\vs)
	+\n\frac{\ep^{\oh\qs}}{\sqrt{\qs}}-\n\sqrt{\ep}
	+\frac{3}{2}\k\frac{\ep^{-2\vs}}{\sqrt{\vs}}
	\Bigr)
	+
	\sqrt{\frac{\alpha_{\QQ}}{\boldsymbol{\sigma}_{\QQ}}+\frac{\g^2\s}{\boldsymbol{\sigma}_{\QQ}^2}
	\Bigl({\cal Q}(\qs)-\oh{\cal Q}(\vs)
	+\n\frac{\ep^{\oh\qs}}{\sqrt{\qs}}-\n\sqrt{\ep}
	+\frac{3}{2}\k\frac{\ep^{-2\vs}}{\sqrt{\vs}}	\Bigr)^2}
	\,.
\end{equation}
For our parameter values, $l_{\Qq}\approx 0.203\,\text{fm}$. This value is rather close to the value of $0.219\,\text{fm}$ obtained for $l_{\Qq}$ in the $Q\bar Qq\bar q$ system \cite{a-QQbqqb}.

The second scale is related to the process of string junction annihilation which occurs at the level of the first excited state. More specifically, it goes like this: $QQqq\bar q\rightarrow Q\bar q+Qqq$, if $\ell$ is increased. In this case we define a critical separation distance by  

\begin{equation}\label{lQ2q3}
	E_{\Q2q3}(\boldsymbol{\ell}_{\Q2q3})=E_{\Qqb}+E_{\Qqq}
	\,.
	\end{equation}
It implies that at the first excited level the system can be thought of mainly as a compact pentaquark if $\ell\leq\boldsymbol{\ell}_{\Q2q3}$. As seen from Figure \ref{all-L}, $\boldsymbol{\ell}_{\Q2q3}$ is of order $0.2\,\text{fm}$. For this range of $\ell$ values, the function $E_{\Q2q3}(\ell)$ is well approximated by Eq.\eqref{penta-factor}. Then a simple calculation shows that 

\begin{equation}\label{lQ2q3-small}
	\boldsymbol{\ell}_{\Q2q3}
=
\frac{\g\sqrt{\s}}{2\boldsymbol{\sigma}_{\QQ}}\Bigl({\cal Q}(\vs)-3\k\frac{\ep^{-2\vs}}{\sqrt{\vs}}\Bigr)
	+
	\sqrt{\frac{\alpha_{\QQ}}{\boldsymbol{\sigma}_{\QQ}}+\frac{\g^2\s}{4\boldsymbol{\sigma}_{\QQ}^2}\Bigl({\cal Q}(\vs)-3\k\frac{\ep^{-2\vs}}{\sqrt{\vs}}\Bigr)^2}
	\,,
\end{equation}
where in the last step we used \eqref{Qqb} and \eqref{EQqq}. One important point is that $\boldsymbol{\ell}_{\Q2q3}$ is finite and scheme independent ($c$ drops out). The other point is that it depends on $\vs$, which describes the position of the baryon vertices in the bulk, rather than on $\qs$ which describes the position of the light quarks. This suggests that $\boldsymbol{\ell}_{\Q2q3}$ is related to gluonic degrees of freedom, as expected from annihilation of string junctions made of gluons. A simple calculation shows that

\begin{equation}\label{lQ2q3-est}
\boldsymbol{\ell}_{\Q2q3}\approx 0.184\,\text{fm}	
\,.
\end{equation}
This value is equal to the value obtained for the $QQ\bar q\bar q$ system in \cite{a-QQqq} and only slightly below the value $\boldsymbol{\ell}_{\2Qq}\approx 0.196\,\text{fm}$ for the $Q\bar Qq\bar q$ system \cite{a-QQbqqb}. Thus, our estimate provides further evidence that in systems with two heavy quark sources the process of string junction annihilation takes place at relatively small heavy quark separations, of order $0.2\,\text{fm}$. Of course the question arises of whether this scale is universal. We will return to this question in Sec.V, after gaining some information about the screening lengths.

The third scale is associated to the transition: $QQq+q\bar q\rightarrow 2Q\bar q+qqq$ which also occurs at the level of the first excited state. Such a transition can be subdivided into two elementary ones, as those of Figure \ref{sint}. The first is the process of string reconnection: $QQq+q\bar q\rightarrow Qqq+Q\bar q$, and the second is the process of string breaking: $Qqq+Q\bar q\rightarrow 2Q\bar q+qqq$. The corresponding equation for a critical distance reads

\begin{equation}\label{lcomp}
	E_{\QQq}(\ell_{\QQq}^{(\qqb)})+E_{\qqb}=2E_{\Qqb}+E_{\nucl}
	\,.
	\end{equation}
Since this scale is large enough, as seen in Figure \ref{all-L}, the solution can be found just by using the asymptotic expansion \eqref{EQQq-large}. So,   

\begin{equation}\label{lcomp2}
\ell_{\QQq}^{(\qqb)}
 =\frac{2}{\ep\sqrt{\s}}
\biggl(
{\cal Q}(\qs)
+
\n\Bigl(\frac{\ep^{\oh \qs}}{\sqrt{\qs}}-\ep^{\oh}\Bigr)
+
\frac{3}{2\sqrt{\qsn}}\Bigl(\k\ep^{-2\qsn}+\n\ep^{\oh\qsn}\Bigr)
+
{\cal I}_{\QQq}
\biggr)
\,,
\end{equation}
where we have used Eqs. \eqref{Eqqb}, \eqref{Qqb} and \eqref{nucl2}. A numerical calculation shows that $\ell_{\QQq}^{(\qqb)}\approx 1.22\,\text{fm}$.  This value is surprisingly close to that found for the string breaking distance in the $Q\bar Q$ system \cite{bulava}.

\section{More Detail On the Potentials}
\renewcommand{\theequation}{4.\arabic{equation}}
\setcounter{equation}{0}

\subsection{An issue with $E_{\qqb}$ and $E_{\nucl}$}

With the help of \eqref{Eqqb} and \eqref{nucl2}, it is straightforward to estimate the rest energies of the pion and nucleon. Using the parameter values of Sec.III, one finds that $E_{\qqb}=1.190\,\text{GeV}$ and $E_{\nucl}=1.769\,\text{GeV}$. These values are considerably larger than the values of $280\,\text{MeV}$ and $1060\,\text{MeV}$ used in the lattice calculations \cite{bulava}.\footnote{This is a reasonable value of the nucleon mass in the case of two flavors because $E_{\nucl}=1060\,\text{MeV}$ at $E_{\qqb}=285\,\text{MeV}$ \cite{nucl}. } The problem is that the current model is not suitable for describing light hadrons as it was originally developed for applications in the heavy quark limit. In the context of string theory on AdS-like geometries, this means that at least one quark is needed to be placed on the boundary of five-dimensional space. 

\begin{figure}[htbp]
\centering
\includegraphics[width=9.25cm]{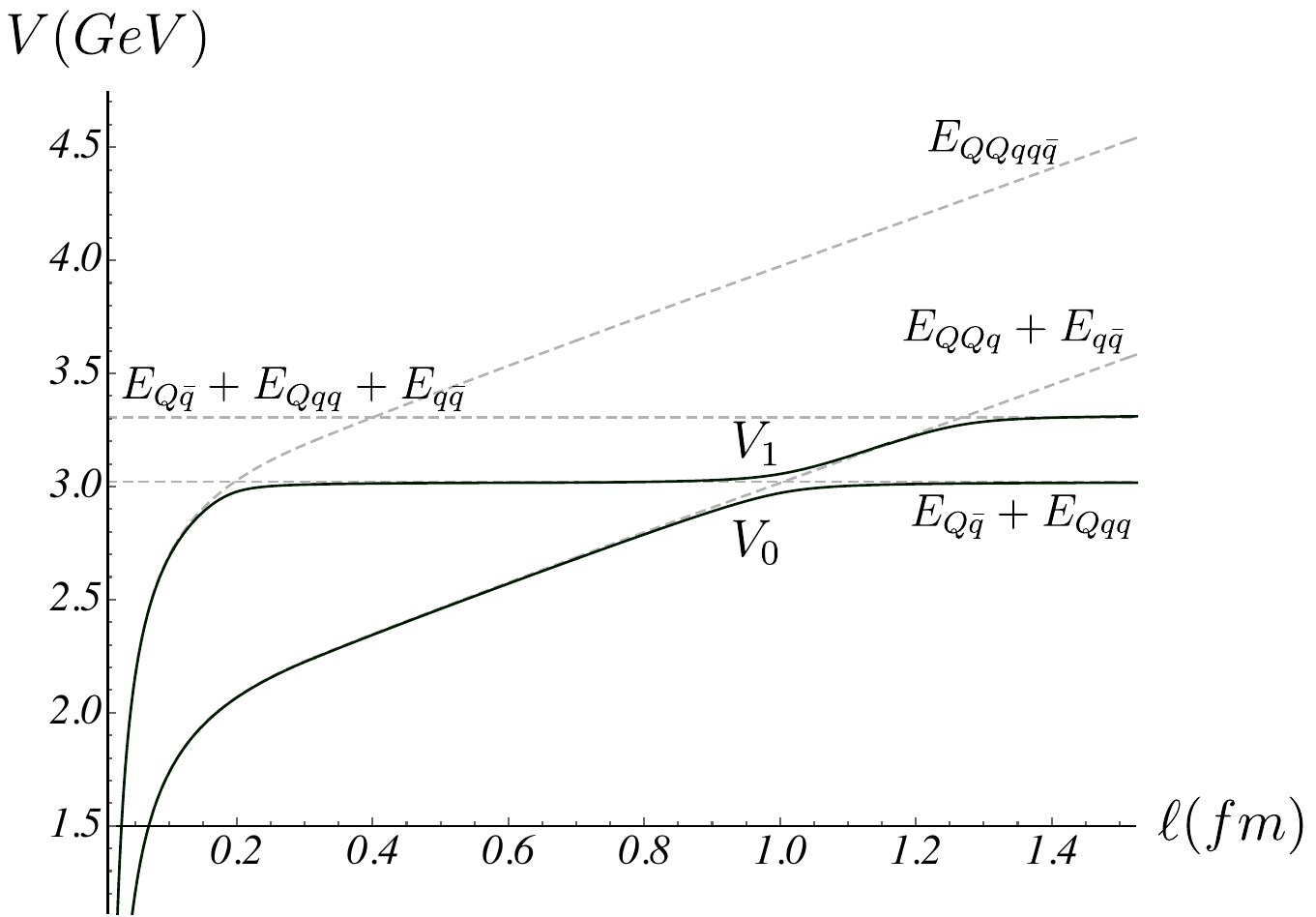}
\caption{{\small Sketched here are the two low-lying B-O potentials of the $QQqq\bar q$ system. The dashed lines indicate the relevant $E$'s.}}. 
\label{V01}
\end{figure}

A possible way out is to treat $E_{\qqb}$ and $E_{\nucl}$ as model parameters. For $E_{\qqb}=280\,\text{MeV}$, the corresponding $E$'s are plotted in Figure \ref{V01}. There are two main differences with the plots of Figure \ref{all-L}. First, the scale of $l_{\Qq}$ is now large, about $1\,\text{fm}$. Second, configuration (e), rather than (f), is relevant for determining the potential $V_1$. So, $V_1=\min\{E_{\Q2q3},E_{\Qqb}+E_{\Qqq}, E_{\QQq}+E_{\qqb},E_{\Qqb}+E_{\Qqq}+E_{\qqb}\}$. Note that there is no change in the behavior of $E_{\Q2q3}$ and $E_{\Qqb}+E_{\Qqq}$. Thus, the value of $\boldsymbol{\ell}_{\Q2q3}$ remains the same and the compact pentaquark is expected at the level of the first excited state.

Since $l_{\Qq}$ is large, one can use the large $\ell$ asymptotic for $E_{\QQq}$ to solve Eq. \eqref{lQqb}. With the help of \eqref{EQQq-large}, one finds that 

\begin{equation}\label{lQq-large}
l_{\Qq}=\frac{3}{\ep\sqrt{\s}}
\biggl(
{\cal Q}(\qs)-\frac{1}{3}{\cal Q}(\vs)
+
\k\frac{\ep^{-2\vs}}{\sqrt{\vs}}
+
\n\frac{\ep^{\oh\qs}}{\sqrt{\qs}}
+
\frac{2}{3}I_{\QQq}-\frac{E_{\qqb}}{3\g\sqrt{\s}}\biggr)
		\,.
\end{equation}
For $E_{\qqb}=280\,\text{MeV}$, $l_{\Qq}\approx 1.001\,\text{fm}$. Interestingly, this value differs only by $3.7\%$ from the value obtained within the $Q\bar Qq\bar q$ system \cite{a-QQbqqb}, where the string reconnection process goes like this: $Q\bar Q+q\bar q\rightarrow Q\bar q+q\bar Q$.

Because of the relevance of configuration (e), the third scale is now related to the process of string breaking: $QQq+q\bar q\rightarrow Q\bar q+Qqq+q\bar q$, which obviously reduces to the process in the $QQq$ system, namely $QQq\rightarrow Q\bar q+Qqq$. The corresponding scale is given by \eqref{lcQQq}. Numerically, $\ell_{\QQq} =1.257\,\text{fm}$ \cite{a-QQq}.

Just like in the case of the $Q\bar Qq\bar q$ system, our results indicate that the following scale hierarchy is met 

\begin{equation}\label{scales}
\boldsymbol{\ell}_{\Q2q3}< l_{\Qq}< \ell_{\QQq}
\,.
\end{equation}
In other words, the scale of string junction annihilation is the smallest and that of string breaking is the largest. One may ask what happens at the physical value of the pion mass. We expect that $\boldsymbol{\ell}_{\Q2q3}$ remains the smallest so that $\boldsymbol{\ell}_{\Q2q3}< l_{\Qq}$, $\ell_{\QQq}$. Of course, this is not the whole answer to this question, but it's certainly an important piece of it.

Having understood the relevant configurations, one can gain some insight into the two low-lying B-O potentials. A natural way for doing so is to consider a model Hamiltonian similar to that used in lattice QCD to study the phenomenon of string breaking \cite{FK}. For the problem at hand it is  

\begin{equation}\label{HV0V1}
{\cal H}(\ell)=
\begin{pmatrix}
\,E_{\QQq}(\ell)+E_{\qqb} & {} & {} & {} & {}\, \\
\,{} & E_{\Qqb}+E_{\Qqq} & {} & {} & \Theta_{ij} \,\\
\,\Theta_{ij}& {} & E_{\Q2q3}(\ell) & {} & {}\, \\
\,{} & {} & {} & {} & E_{\Qqb}+E_{\Qqq}+E_{\qqb}\,\\
\end{pmatrix}
\,,
\end{equation}
where the off-diagonal elements describe the strength of mixing between the four states (string configurations). The potentials of interest are the two smallest eigenvalues of the matrix ${\cal H}$. 

Unlike lattice QCD, where the Hamiltonian could in principle be extracted from a correlation matrix, it is not clear how to compute the off-diagonal elements within the string models. This makes it difficult to see precisely how the potentials look like. Nevertheless, we can learn from our experience with the other quark systems about the order of magnitude of the $\Theta$'s near the transition points.\footnote{For example, one can take these $\Theta$'s to be of order $47\,\text{MeV}$, as in the case of the $Q\bar Q$ system \cite{bulava}.} With the help of this, the picture will then look more like what is shown in Figure \ref{V01}. The compact pentaquark configuration contributes dominantly to the potential $V_1$ at heavy quark separations smaller than $0.2\,\text{fm}$.
\subsection{Simple approximations to $V_1$}

For practical purposes, the parametric expressions for the potentials may look somewhat cumbersome. On the other hand, a simple parametrization motivated by lattice calculations was proposed for the $V_0$ potential of the $QQ\bar q\bar q$ system in \cite{wagner}. From a string point of view the description includes two distinct configurations, one associated with a compact tetraquark and the other with a pair of non-interacting mesons \cite{a-QQqq}. This is similar to what we have seen above for $V_1$. Indeed, at separations $\ell\lesssim 0.6\,\text{fm}$, this potential is determined in terms of the two string configurations, one of which corresponds to the compact pentaquark. The physical reason for this similarity is, in fact, the underlying heavy quark-diquark symmetry.\footnote{We return to this in Sect.V.}  

Thus, it is reasonable, following \cite{wagner}, to suggest that for $\ell\lesssim 0.6\,\text{fm}$

\begin{equation}\label{latticeV1}
V_1=-\frac{\alpha}{\ell}\exp\Bigl\{-\frac{\ell^{\,p}}{d^{\, p}}\,\Bigl\}\,+E_{\Qqb}+E_{\Qqq}
\,,
\end{equation}
with parameters $\alpha$, $d$, and $p$. To express these parameters in terms of 
ours, we use the small $\ell$ expansion \eqref{penta-factor} and solve for the unknown coefficients. As a result, except for the constant term, we get

\begin{equation}\label{dp}
	\alpha=\alpha_{\QQ}
	\,,\qquad
	d=\sqrt{\frac{\alpha_{\QQ}}{\boldsymbol{\sigma}_{\QQ}}}
	\,,\qquad
	p=2
	\,.
\end{equation}
It is no coincidence that they are the same as those of the $QQ\bar q\bar q$ \cite{a-QQqq} system. This is a consequence of heavy quark-diquark symmetry. In particular, a numerical calculation gives for the screening length $d\approx 0.200\,\text{fm}$. 

\begin{figure}[htbp]
\centering
\includegraphics[width=8.25cm]{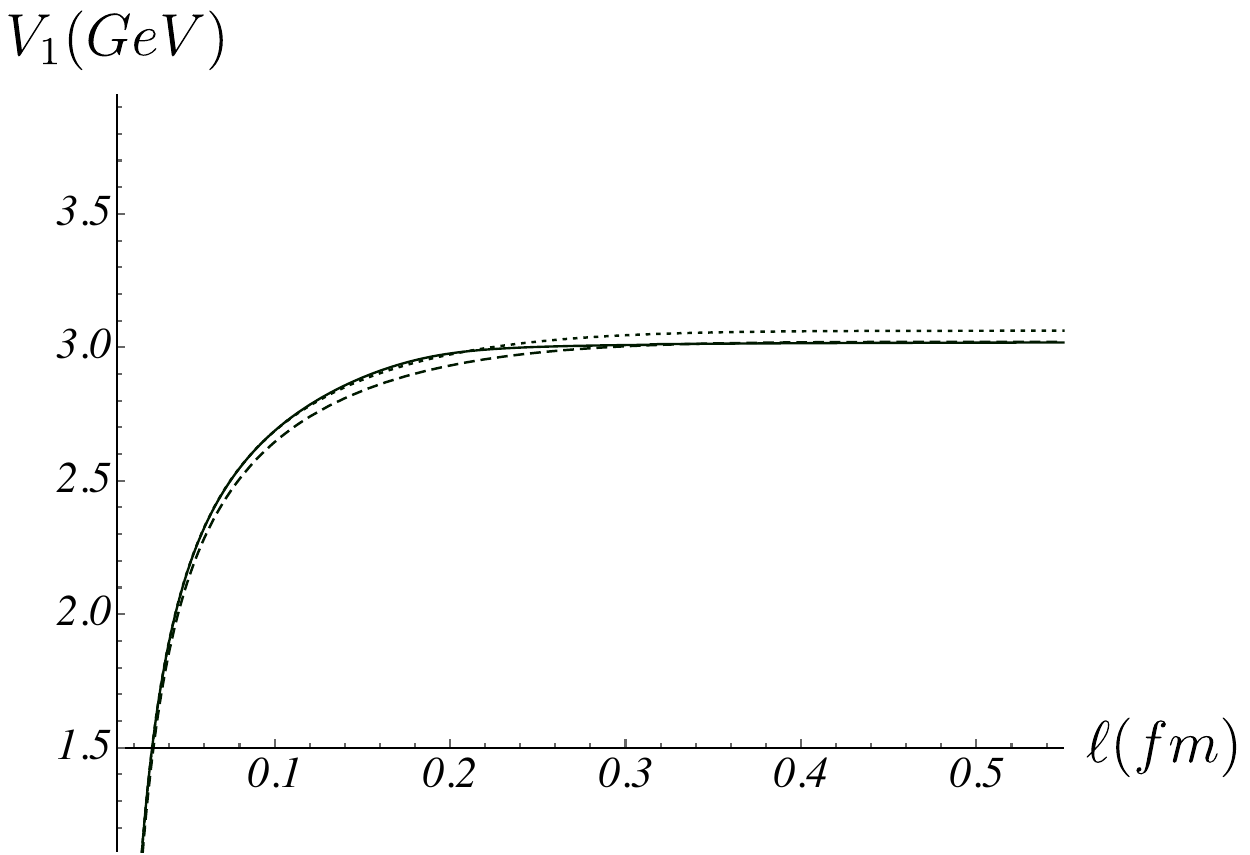}
\caption{{\small $V_1$ vs $\ell$ for $\ell\lesssim 0.6\,\text{fm}$. The solid curve corresponds to \eqref{V1-small}, whereas the dashed and dotted curves to the approximations \eqref{latticeV1} and \eqref{latticeV1UV}.}}
\label{V1s}
\end{figure}

For the given range of $\ell$, the model Hamiltonian is well approximated by a $2\times 2$ submatrix of ${\cal H}$. Its diagonal elements are given by $E_{\Q2q3}$ and $E_{\Qqb}+E_{\Qqq}$, and an off-diagonal element by $\Theta_{\Q2q3}$. This leads to the following expression for $V_1$: 

\begin{equation}\label{V1-small}
V_1=\oh\Bigl(E_{\Q2q3}+E_{\Qqb}+E_{\Qqq}\Bigr)
-
\sqrt{\frac{1}{4}\Bigl(E_{\Q2q3}-E_{\Qqb}-E_{\Qqq}\Bigr)^2+\Theta_{\Q2q3}^2}
\,.	
\end{equation}
It is now interesting to compare it to the approximation \eqref{latticeV1}. Figure \ref{V1s} shows the graphs for both cases. We set 
$\Theta_{\Q2q3}=47\,\text{MeV}\,\text{GeV}$ to have a small deviation between the curves. 

There are two observations to be made here. The first is a falloff at large $\ell$. It is power-law for \eqref{V1-small}, but exponential for \eqref{latticeV1}.\footnote{The difference between these is slightly visible around $\ell=0.3\,\text{fm}$.} The reason for the power-law falloff is the choice of $\Theta=const$. In fact, one can get the exponential falloff by taking $\Theta$ as a Gaussian function, but then one additional parameter is required, a Gaussian width. The second is a more visible deviation between the solid and dashed curves in the interval $0.05\,\text{fm}\lesssim \ell\lesssim 0.23\,\text{fm}$. The appearance of this deviation can be ascribed to the mismatch of the constant terms in the small $\ell$ expansions of \eqref{latticeV1} and \eqref{V1-small}. The point is that the constant term dictated by quark-diquark symmetry for \eqref{latticeV1} is equal to $c+E_{\Qqbqq}$, with $c$ coming from the heavy quark-quark potential in \eqref{EQQq-small}. If so, then the improved approximation to $V_1$ takes the form 

\begin{equation}\label{latticeV1UV}
V_1=-\frac{\alpha_{\QQ}}{\ell}
\exp\Bigl\{-\frac{\boldsymbol{\sigma}_{\QQ}}{\alpha_{\QQ}}\,\ell^2\Bigl\}\,+c+E_{\Qqbqq}
\,.
\end{equation}
As seen from the Figure, now there is no visible deviation in the above interval, but instead another deviation comes out for larger $\ell$.  This is so because of the difference between the constant terms in \eqref{latticeV1} and \eqref{latticeV1UV}.\footnote{Numerically, $E_{\Qqbqq}+c-E_{\Qqb}-E_{\QQq}\approx 41\,\text{MeV}$.} 

The lesson to learn from this example is that a consistent parametrization should be based on a function having different constant terms in its small and large $\ell$ expansions, as for example in \eqref{V1-small}. Hopefully it will be possible eventually to determine $V_1$ reliably by computer simulations.

\subsection{Another view on the $QQqq\bar q$ system}

According to one of our assumptions made in Sec.II, the string configurations corresponding to excited states are constructed by adding $q\bar q$ pairs to the basic configurations corresponding to the ground state. This implies that the $QQqq\bar q$ system can be thought of as a subsystem of the $QQq$ system obtained by adding one pair.\footnote{We did not adopt this viewpoint in Sec.III, in part because it might have made the analysis more tedious.} Here we briefly explore such an idea using the results of the previous Section and Appendix B. This will also enable us to show some of the subtleties of the $QQq$ system and thereby extend the analysis of Ref.\cite{a-QQq}.

We start with the ground state energy. As discussed in Appendix B, from the string theory point of view, it is described in terms of the configurations of Figure \ref{4QQq} and the corresponding potential is defined by $V_0=\min\{E_{\QQq}, E_{\Qqb}+E_{\Qqq}\}$. The transition between these two occurs because of string breaking. The potential, which is the same as in Figure \ref{VQQq}, is shown in Figure \ref{V012} below.\footnote{Like in Figure \ref{V01}, the resulting plots of the $V_i$'s require a caveat. Here too, we are not aware precisely of any $\Theta_{ij}$.}
\begin{figure}[htbp]
\centering
\includegraphics[width=9.6cm]{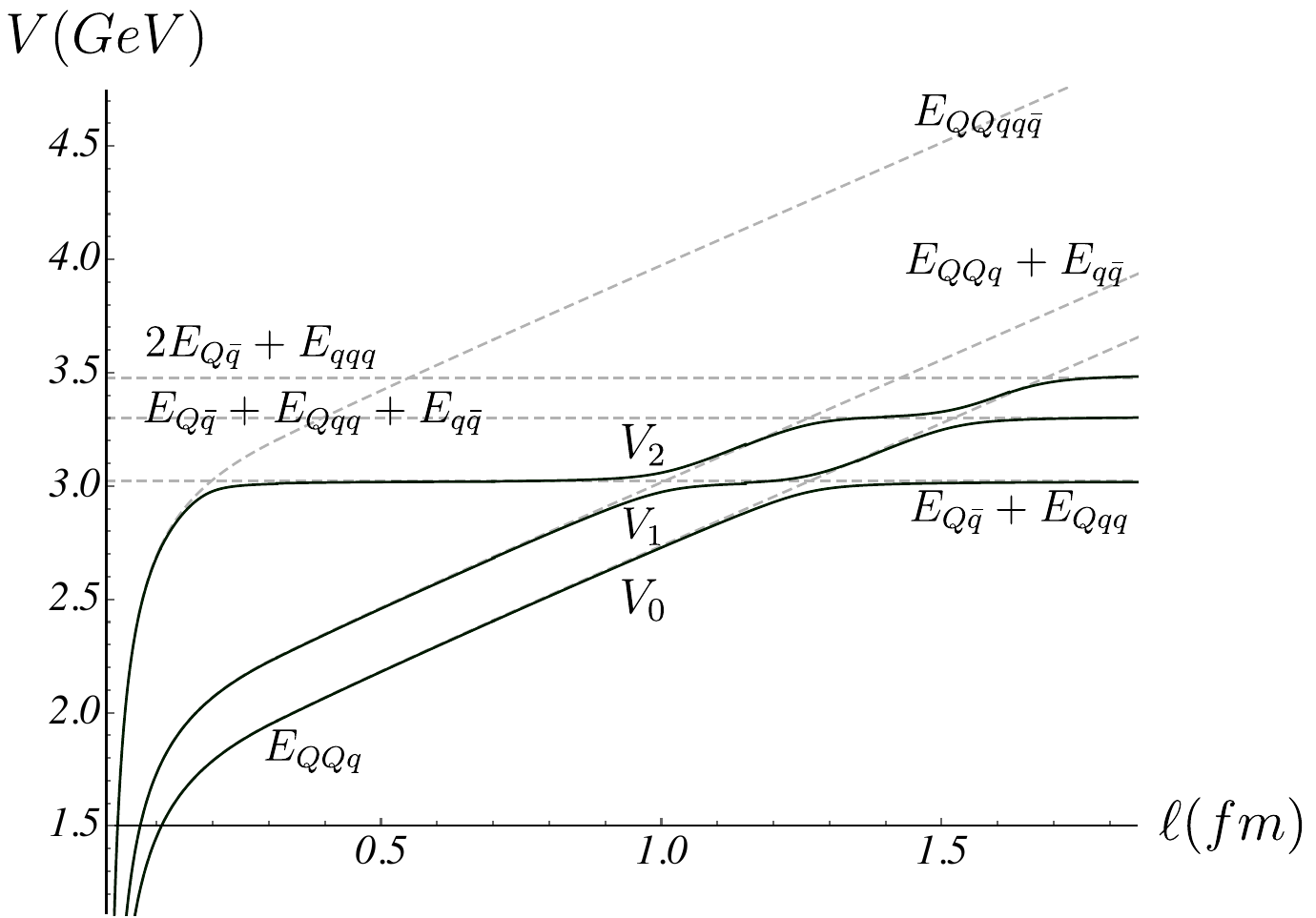}
\caption{{\small Sketched here are the three low-lying B-O potentials of the $QQq$ system. Various $E$ vs $\ell$ plots are indicated by dashed curves and horizontal lines. Here $E_{\qqb}=280\,\text{MeV}$ and $E_{\nucl}=1060\,\text{MeV}$.}}
\label{V012}
\end{figure}

The energy of the first excited state is described in terms of four configurations. These are the configurations of Figures \ref{c40}, \ref{c42}(e), and \ref{4QQq}(a). So, we have $V_1=\min\{E_{\QQq}+E_{\qqb}, E_{\Qqb}+E_{\Qqq}, E_{\QQq},E_{\Qqb}+E_{\Qqq}+E_{\qqb}\}$. As seen from the Figure, for $\ell\lesssim 1.15\,\text{fm}$ the potential looks similar to what is shown in Figure \ref{V01} for the ground state energy. Thus, the transition at $\ell=1.001\,\text{fm}$ is due to string reconnection, with the formula \eqref{lQq-large} remaining valid. For larger separations there is no similarity between them. The transition at $\ell=1.257\,\text{fm}$ is now interpreted as a result of fusion of two strings: $Q\bar q+Qqq\rightarrow QQq$. This process is inverse to the process of string breaking, which is discussed in Appendix B. Because of this, the formula \eqref{lcQQq} holds true. The final transition near $\ell =1.51 \,\text{fm}$ is an example of double string breaking: $QQq\rightarrow Q\bar q+Qqq+q\bar q$, as sketched in Figure \ref{sint}(b').\footnote{Another possibility could be that two different strings break down. One of which is attached to a heavy quark, and the other to a light quark. However, from the four dimensional perspective this doesn't seem possible because the light quark sits on top of a string junction so that the corresponding string is shrunk into a point.} By analogy with what we did for the other transitions, we define a critical separation distance by  

\begin{equation}\label{lcdouble}
	E_{\QQq}(\ell_{\QQq}^{(2)})=E_{\Qqb}+E_{\Qqq}+E_{\qqb}
	\,.
	\end{equation}
For large $\ell$, this equation can be easily solved by using the asymptotic expression \eqref{EQQq-large}. The solution is given by 

\begin{equation}\label{lQQq-double}
\ell_{\QQq}^{(2)}=\frac{3}{\ep\sqrt{\s}}
\biggl(
{\cal Q}(\qs)-\frac{1}{3}{\cal Q}(\vs)
+
\k\frac{\ep^{-2\vs}}{\sqrt{\vs}}
+
\n\frac{\ep^{\oh\qs}}{\sqrt{\qs}}
+\frac{2}{3}I_{\QQq}+\frac{E_{\qqb}}{3\g\sqrt{\s}}\biggr)
		\,.
\end{equation}
A simple calculation shows that for $E_{\qqb}=280\,\text{MeV}$, $\ell_{\QQq}^{(2)}= 1.513 \,\text{fm}$. 

Before proceeding further, we pause here to briefly discuss the process of double string breaking in the $QQq$ and $Q\bar Q$ systems. In the $QQq$ system, we consider a string attached to a heavy quark so that the critical separation distance is given by \eqref{lQQq-double}. It is straightforward to extend the above analysis to the $Q\bar Q$ system. First, one has to consider the process: $Q\bar Q\rightarrow Q\bar q+q\bar Q+q\bar q$, and then to define a critical separation distance by 

\begin{equation}\label{lcQQ2}
	E_{\QQb}(\ell_{\QQb}^{(2)})=2E_{\Qqb}+E_{\qqb}
	\,.
	\end{equation}
Using the large $\ell$ expansion for $E_{\QQb}(\ell)$ of \cite{a-strb}, one gets

\begin{equation}\label{lQQb-double}
\ell_{\QQb}^{(2)}=\frac{2}{\ep\sqrt{\s}}
\biggl(
{\cal Q}(\qs)
+\n\frac{\ep^{\oh\qs}}{\sqrt{\qs}}
+I_{0}
+\frac{E_{\qqb}}{2\g\sqrt{\s}}\biggr)
		\,.
\end{equation}
Here $I_0$ is defined in Appendix A. If, for instance, $E_{\qqb}=280\,\text{MeV}$, then $\ell_{\QQb}^{(2)}= 1.476\,\text{fm}$. This value is very close (within $2.4\%$) to that obtained above for the $QQq$ system. Interestingly, a similar story also holds for the standard string breaking  (by one quark pair) distances, where the difference is about $2.9\%$ \cite{a-QQq}.

The energy of the second excited state is described in terms of six configurations. These are shown in Figures \ref{c40}, \ref{c41}, \ref{c42}(e)-(f), and \ref{4QQq}(a). We therefore have $V_2=\min\{E_{\Q2q3}, E_{\Qqb}+E_{\Qqq}, E_{\QQq}+E_{\qqb},E_{\Qqb}+E_{\Qqq}+E_{\qqb}, E_{\QQq}, 2E_{\Qqb}+E_{\nucl}\}$. As for $V_1$, a quite similar story holds true for $V_2$. For $\ell\lesssim 1.35\,\text{fm}$ it looks similar to what is shown in Figure \ref{V01} for the energy of the first excited state. So, the transition at $\ell=0.184\,\text{fm}$ is due to the process of string junction annihilation, with the formula \eqref{lQ2q3-small} remaining valid. The next transition at $\ell=1.001\,\text{fm}$ is interpreted as a result of string reconnection. In this case, the critical separation distance is given by \eqref{lQq-large}. The transition at $\ell=1.257\,\text{fm}$ is due to string breaking, as discussed in Appendix B. The remaining transition at $\ell=1.513\,\text{fm}$ can be interpreted as a result of fusion of two strings: $Q\bar q+Qqq+q\bar q\rightarrow QQq$. Clearly, it is inverse to the process of double string breaking so that the above formula can also be applied to estimate the critical separation distance. 

So far we have considered the domain of $V_2$, where the analysis is straightforward and relies on the previous results. However, for larger separations a new analysis is needed. The transition near $\ell=1.65\,\text{fm}$ is another example of double string breaking: $QQq\rightarrow 2Q\bar q+qqq$. A crucial difference from the first example is that two strings break down. Both have a heavy quark at their endpoints. For each of them, the process of string breaking is as sketched in Figure \ref{sint}(b). In this case we define a critical separation distance by  

\begin{equation}\label{lcdouble'}
	E_{\QQq}(\ell_{\QQq}^{(2')})=2E_{\Qqb}+E_{\nucl}
	\,.
	\end{equation}
As before, it is easy to find a solution for large $\ell$. It is 
\begin{equation}\label{lQQq-double'}
\ell_{\QQq}^{(2')}=\frac{2}{\ep\sqrt{\s}}
\biggl(
{\cal Q}(\qs)
+\n\frac{\ep^{\oh\qs}}{\sqrt{\qs}}
+I_{\QQq}
+\frac{E_{\nucl}}{2\g\sqrt{\s}}\biggr)
		\,.
\end{equation}
For $E_{\nucl}=1060\,\text{MeV}$, a simple estimate gives $\ell_{\QQq}^{(2')}= 1.661\,\text{fm}$ that is smaller than the value of $\ell_{\QQq}^{(2)}$.

We conclude that in the $QQq$ system the pentaquark configuration makes the dominant contribution to the potential of the second excited state at small separations. This is quite similar to what happen in the $QQqq\bar q$ system to the potential of the first excited state.  

\section{Concluding comments}
\renewcommand{\theequation}{5.\arabic{equation}}
\setcounter{equation}{0}

\begin{figure}[htbp]
\centering
\includegraphics[width=8.25cm]{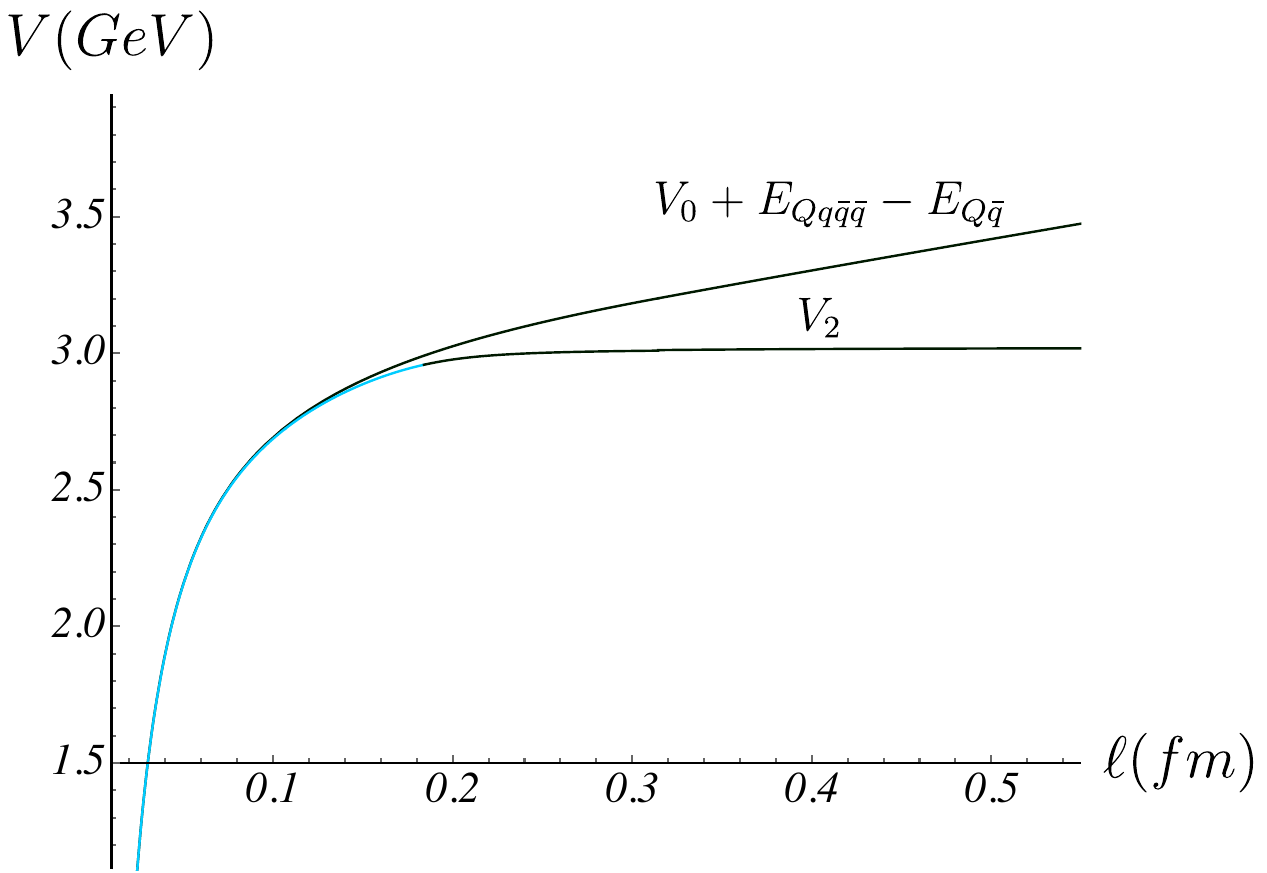}
\caption{{\small The potentials $V_0$, shifted by $E_{\Qqqq}-E_{\Qqb}$, and $V_2$. The part of $V_2$ which is well approximated by $E_{\Q2q3}$ is shown in blue.}}
\label{V02}
\end{figure}

(i) An interesting relation between the energies of some connected configurations can be deduced from heavy quark-diquark symmetry. Indeed, one can express $E_{\QQ}$ from Eq.\eqref{EQQq-small} and then substitute it into Eq.\eqref{penta-factor} to get 

\begin{equation}\label{E-relation}
E_{\Q2q3}(\ell)=E_{\QQq}(\ell)+E_{\Qqqq}-E_{\Qqb}
\,.
\end{equation}
The above derivation holds for relatively small $\ell$. We have developed all the necessary machinery to directly check if this relation holds for larger values of $\ell$. To do so, it is convenient to think in terms of the $QQq$ system and to further restrict discussion to the potentials of the ground and second excited states. As seen from Figure \ref{V012}, $V_0$ can be well approximated by $E_{\QQq}$ up to separations of order $1,2\,\text{fm}$, while $V_2$ by $E_{\Q2q3}$ up to separations of order $0.2\,\text{fm}$. The reason for the latter is the flattening of $V_2$ near $\ell=0.2\,\text{fm}$ due to string junction annihilation. For illustration, in Figure \ref{V02} we plot the potentials $V_2$  
and $V_0$ shifted by the constant $E_{\Qqqq}-E_{\Qqb}$. Obviously, the refined version of \eqref{E-relation} is 

\begin{equation}\label{V-relation}
V_2(\ell)=V_0(\ell)+E_{\Qqqq}-E_{\Qqb}
\,,\qquad
\text{if}
\qquad
\ell\lesssim 0.2\,\text{fm}
\,.
\end{equation}

It is tempting to apply this relation in the heavy quark limit, where contributions from the motion of the heavy quarks and spin interactions are negligible, to derive a relation among the masses of heavy-light and doubly-heavy-light hadrons. We will thus have 

\begin{equation}\label{M-relation}
m_{\Q2q3}-m_{\QQq}=m_{\Qqqq}-m_{\Qqb}
\,.
\end{equation}
It is important to realize, however, that because of the use of \eqref{V-relation}, the doubly heavy hadrons are compact in sense of heavy quark separation.

Similar formulas arise in the case of the $QQ\bar q \bar q$ system \cite{a-QQqq}. Again, the phenomenon of string junction annihilation plays a pivotal role in establishing the upper bound on $\ell$. In the heavy quark limit one can use those to derive the relation among the masses of heavy-light and doubly-heavy-light mesons and baryons as that of \cite{quigg}. Moreover, by essentially the same arguments that we have given above and in \cite{a-QQqq}, one can show that similar relations hold between the corresponding potentials of the $QQqq\bar q$ and $QQ\bar q\bar q$ systems. In this case, the relation among masses reads\footnote{Alternatively, it can be obtained by combining \eqref{M-relation} with the relation of \cite{quigg}.}

\begin{equation}\label{M2-relation}
m_{\Q2q3}-m_{\QQqq}=m_{\Qqqq}-m_{\Qqq}
\,.
\end{equation}

It is interesting to see what results come out of both relations and how those agree or disagree with phenomenological values. For reasons of brevity, we will not discuss all this here, but only make a crude estimate. Consider the first relation. For the mass of the $QQqq\bar q$ state several values can be found in the literature \cite{wang,FG,zhu}. We use these as an input to estimate the mass of $Qq\bar q \bar q$ and compare the output with the result of a direct calculation in \cite{lu}. Our results are presented in Table \ref{estimates}. 
\begin{table*}[htb]
\renewcommand{\arraystretch}{2}
\centering 	
\begin{tabular}{lccccccr}				
\hline
\hline
State ~~~~~&~~~~~~~\cite{wang} ~~~~~&~~~~~\cite{FG} ~~~~~&~~~~~\cite{zhu}~~~~~& ~~~~~\cite{lu}~~~~~
\rule[-3mm]{0mm}{8mm}
\\
\hline 
$ccqq\bar q$ & 4.21  & 4.54 & 4.70 & $-$  \\
$bbqq\bar q$ & 10.75 & 11.15 & 11.37 & $-$ \\
\hline
$cq\bar q\bar q$ & 2.41  & 2.74 & 2.90 & $2.57$ \\
$bq\bar q\bar q$ & 5.85  & 6.25 & 6.47 & $5.98$ \\
\hline
\hline
\end{tabular}
\caption{ \small Masses (in GeV) of $QQqq\bar q$ and $Qq\bar q \bar q$ states.}
\label{estimates}
\end{table*}
The values used here for the masses of $QQq$ are the same as the smallest values in \cite{quigg}, namely $m_{ccq}=3.66\,\text{GeV}$ and $m_{bbq}=10.18\,\text{GeV}$. The masses of $Q\bar q$ are those of the $D^0$ and $B^{\pm}$ mesons \cite{PDG}, $m_{c\bar q}=1.86\,\text{GeV}$ and $m_{b\bar q}=5.28\,\text{GeV}$. Since the ranges of values are quite wide, it is not surprising that both values of \cite{lu} are inside those ranges. While it is surprising that for $cq\bar q\bar q$ all the values don't exceed $2.90\,\text{GeV}$, which is the mass of the $X_0(2900)$ state whose flavor content is $ud\bar s\bar c$. 

\begin{figure}[htbp]
\centering
\includegraphics[width=14cm]{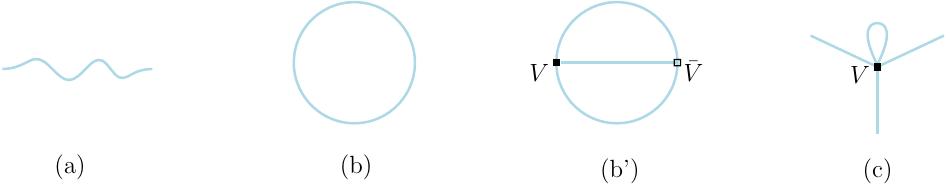}
\caption{{\small Some gluonic excitations.}}
\label{excitations}
\end{figure}

(ii) The assumptions made in Sect.II are oversimplified for various reasons. For one thing, excited strings, as one sketched in Figure \ref{excitations}(a), must be included when building string configurations for excited states. These strings 
represent a kind of gluonic excitations of the system which have been extensively studied in the literature, but only within the $Q\bar Q$ system. Also, glueballs must be included. They are another kind of gluonic excitations. The two simplest examples are sketched in Figures \ref{excitations}(b) and (b'). The first is just a closed string, and the second involves a pair of vertices connected by strings. Both kinds of gluonic excitations are natural from the point of view of string theory in four dimensions \cite{XA}. There is however a novelty related to the description of the baryon vertex as a five-brane in ten dimensions \cite{witten}. Accordingly, brane excitations must be included. This would lead to a bunch of excited vertices. One example of these is sketched in Figure  \ref{excitations} (c), where excitation is due to an open string with endpoints on the brane. One can think of this as a new kind of gluonic excitations. It would be of great interest to find evidence for this kind of excitations by means of lattice simulations.

(iii) As we have seen, at small heavy quark separations the pentaquark configuration makes the dominant contribution to the first excited B-O potential of the $QQqq\bar q$ system. This does not exclude the possibility of its dominance at larger separations, but for higher excited potentials. The key difference with the first case is that if one thinks in terms of the heavy quark separation, the pentaquarks are compact (small) in the first case but not in the others.   

(iv) An interesting observation, which one can make from Eq.\eqref{lQ2q3-small} and Eq.(3.42) in \cite{a-QQqq}, is that in the $QQqq\bar q$ and $QQ\bar q\bar q$ systems the critical separation distances characterizing the process of string junction annihilation are equal to each other. So, 

\begin{equation}\label{sja-scale}
\boldsymbol{\ell}_{\Q2q3}=\boldsymbol{\ell}_{\QQqq}
\,.	
\end{equation}
This is also true for the screening lengths, if one uses the Gaussian form for the corresponding potentials (see Eq.\eqref{dp} and Eq.(4.5) in \cite{a-QQqq}). At first glance, the reason for this is that both systems possess heavy quark-diquark symmetry. It is sufficient for guaranteeing the equal screening lengths, but not the equal critical separation distances. There one needs some peculiar relation among the rest energies of heavy-light hadrons. So we are left with the two important questions: 1) Is the critical separation distance $\boldsymbol{\ell}$ universal for doubly heavy quark systems? 2) Why is it so small? Hopefully, lattice QCD will be able to make further progress in addressing these questions in the near future.
\begin{acknowledgments}
We would like to thank I.Ya. Aref'eva, J.-M. Richard, and M. Wagner for useful communications and conversations. This work was supported in part by Russian Science Foundation grant 20-12-00200 in association with Steklov Mathematical Institute.

\end{acknowledgments}

\appendix

\section{Notation}
\renewcommand{\theequation}{A.\arabic{equation}}
\setcounter{equation}{0}
In all Figures throughout the paper, heavy and light quarks (antiquarks) are denoted by $Q$ and $q\,(\bar q)$, and baryon (antibaryon) vertices by $V\,(\bar V)$. Straight lines represent non-excited strings. Light quarks (antiquarks) are set at $r=\rq\,(\rqb)$ and vertices at $r=\rv\,(\rvb)$, unless otherwise indicated. It is convenient to introduce dimensionless variables: $q=\s\rq^2$, $\bar q=\s\rqb^2$, $v=\s\rv^2$, and $\bar v=\s\rvb^2$. They take values between $0$ and $1$, and show how far from the soft-wall these objects are.\footnote{The soft wall in such units is located at $1$.} To classify the critical separations related to the string interactions of Figure \ref{sint}, the notation $l$ is used for (a), $\ell$ for (b-b'), and $\boldsymbol{\ell}$ for (c).

To present the resulting formulas in a compact form, we use the set of basic functions \cite{a-stb3q}: 

\begin{equation}\label{fL+}
{\cal L}^+(\alpha,x)=\cos\alpha\sqrt{x}\int^1_0 du\, u^2\, \ep^{x (1-u^2)}
\Bigl[1-\cos^2{}\hspace{-1mm}\alpha\, u^4\ep^{2x(1-u^2)}\Bigr]^{-\frac{1}{2}}
\,,
\qquad
0\leq\alpha\leq\frac{\pi}{2}\,,
\qquad 
0\leq x\leq 1
\,.
\end{equation}
${\cal L}^+$ is a non-negative function which vanishes if $\alpha=\frac{\pi}{2}$ or $x=0$, and has a singular point at $(0,1)$;

\begin{equation}\label{fL-}
{\cal L}^-(y,x)=\sqrt{y}
\biggl(\,
\int^1_0 du\, u^2\, \ep^{y(1-u^2)}
\Bigl[1-u^4\,\ep^{2y(1-u^2)}\Bigr]^{-\frac{1}{2}}
+
\int^1_
{\sqrt{\frac{x}{y}}} 
du\, u^2\, \ep^{y(1-u^2)}
\Bigl[1-u^4\,\ep^{2y(1-u^2)}\Bigr]^{-\frac{1}{2}}
\,\biggr)
\,,
\quad
0\leq x\leq y\leq 1
\,.
\end{equation}
${\cal L}^-$ is also a non-negative function. It vanishes at the origin and becomes singular at $y=1$. There is a simple relation between ${\cal L}^+$ and ${\cal L}^-$, namely ${\cal L}^+(0,x)={\cal L}^-(x,x)$;

\begin{equation}\label{fE+}
{\cal E}^+(\alpha,x)=\frac{1}{\sqrt{x}}
\int^1_0\,\frac{du}{u^2}\,\biggl(\ep^{x u^2}
\Bigl[
1-\cos^2{}\hspace{-1mm}\alpha\,u^4\ep^{2x (1-u^2)}
\Bigr]^{-\frac{1}{2}}-1-u^2\biggr)
\,,
\qquad
0\leq\alpha\leq\frac{\pi}{2}\,,
\qquad 
0\leq x\leq 1
\,.
\end{equation}
${\cal E}^+$ is singular at $x=0$ and $(0,1)$;

\begin{equation}\label{fE-}
{\cal E}^-(y,x)=\frac{1}{\sqrt{y}}
\biggl(
\int^1_0\,\frac{du}{u^2}\,
\Bigl(\ep^{y u^2}\Bigl[1-u^4\,\ep^{2y(1-u^2)}\Bigr]^{-\frac{1}{2}}
-1-u^2\Bigr)
+
\int^1_{\sqrt{\frac{x}{y}}}\,\frac{du}{u^2}\,\ep^{y u^2}
\Bigl[1-u^4\,\ep^{2y(1-u^2)}\Bigr]^{-\frac{1}{2}}
\biggr) 
\,,
\,\,\,
0\leq x\leq y\leq 1
\,.
\end{equation}
${\cal E}^-$ is singular at $(0,0)$ and $y=1$. Just like for the ${\cal L}$'s, one also has ${\cal E}^+(0,x)={\cal E}^-(x,x)$;

\begin{equation}\label{Q}
{\cal Q}(x)=\sqrt{\pi}\text{erfi}(\sqrt{x})-\frac{\ep^x}{\sqrt{x}}
\,,
\end{equation}
which is the special case of ${\cal E}^+$ with $\alpha=\frac{\pi}{2}$. Here $\text{erfi}(x)$ is the imaginary error function. A useful fact is that its small-$x$ behavior is given by 

\begin{equation}\label{Q0}
{\cal Q}(x)=-\frac{1}{\sqrt{x}}+\sqrt{x}+O(x^{\frac{3}{2}})
\,;
\end{equation}

\begin{equation}\label{I}
	{\cal I}(x)=
	I_0
	-
	\int_{\sqrt{x}}^1\frac{du}{u^2}\ep^{u^2}\Bigl[1-u^4\ep^{2(1-u^2)}\Bigr]^{\frac{1}{2}}
	\,,
\qquad
I_0=\int_0^1\frac{du}{u^2}\Bigl(1+u^2-\ep^{u^2}\Bigl[1-u^4\ep^{2(1-u^2)}\Bigr]^{\frac{1}{2}}\Bigr)
\,,
\qquad
0< x\leq 1
\,.
\end{equation}
Numerically, $I_0=0.751$.

\section{The $QQq$ system}
\renewcommand{\theequation}{B.\arabic{equation}}
\setcounter{equation}{0}

We will give a brief review of the string construction for the $QQq$ system proposed in \cite{a-QQq}, whose conventions we follow. We are limited here to the potential $V_0$ corresponding to the ground state.
 
\begin{figure}[htbp]
\centering
\includegraphics[width=9.25cm]{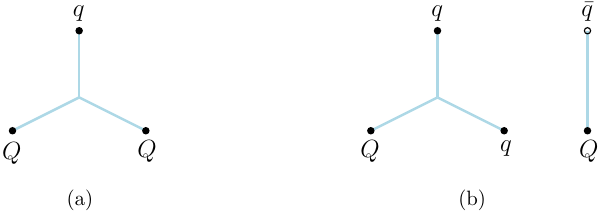}
\caption{{\small String configurations contributing to the potential $V_0$ of the $QQq$ system.}}
\label{4QQq}
\end{figure}

From the point of view of four dimensional string models \cite{XA}, the only relevant string configurations are those shown in Figure \ref{4QQq}. 
The first consists of the valence quarks joined by strings. The strings meet at the string junction. This is a usual string picture for doubly heavy baryons. The second is obtained by adding a virtual pair $q\bar q$ to the first configuration. It is the same as that of Figure \ref{c40} (b) which describes  
two non-interacting hadrons: $Qqq$ and $Q\bar q$.\footnote{Another disconnected configuration consisting of $QQq$ and $q\bar q$ doesn't contribute to the ground state.} The reason that the second configuration contributes to the ground state is that for large $\ell$ its energy is of order $1$, whereas the energy of the first configuration is of order $\ell$. The transition between the two different regimes corresponds to the baryon decay  

\begin{equation}\label{decayQQq}
QQq\rightarrow Qqq\,+\,Q\bar q
\,.
\end{equation}
In the language of string theory, such a decay can be interpreted as string breaking. One of the strings attached to the heavy quarks breaks down.  

\begin{figure}[htbp]
\centering
\includegraphics[width=4.2cm]{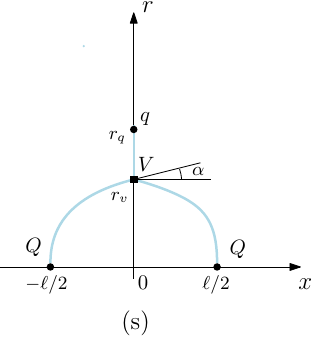}
\hspace{0.7cm}
\includegraphics[width=5.75cm]{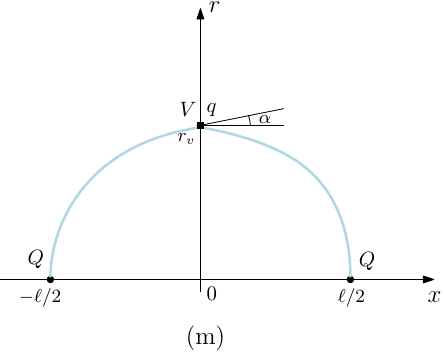}
\hspace{0.7cm}
\includegraphics[width=5.95cm]{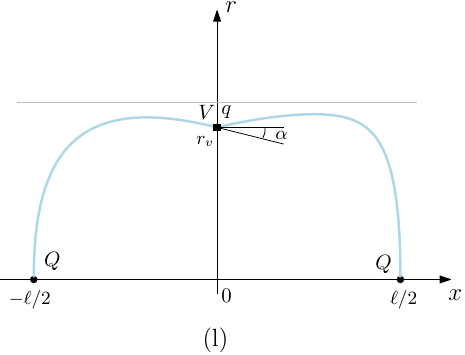}
\caption{{\small Three types of connected configurations for the $QQq$ system in five dimensions. $\alpha$ denotes the tangent angle of the left string. The gray horizontal line in (l) represents the soft wall at $r=1/\sqrt{\s}$.}}
\label{cQQq}
\end{figure}

Now consider these configurations in the five-dimensional framework. We begin with the connected configuration of Figure \ref{4QQq}(a). The important point here is that the shape of the configuration changes with the increase of heavy quark separation. As a result, the single string configuration in four dimensions is replaced by three different configurations in five dimensions, as those of Figure \ref{cQQq}. 

For small $\ell$ the corresponding configuration is labeled by (s). In this 
case, the total action is the sum of the Nambu-Goto actions for the three fundamental strings plus the actions for the baryon vertex and background scalar. The relation between the energy and heavy quark separation is written in parametric form 

\begin{equation}\label{EQQqs}
\ell=\frac{2}{\sqrt{\s}}{\cal L}^+(\alpha,v)
\,,\qquad
E_{\QQq}=\g\sqrt{\s}
\Bigl(
2{\cal E}^+(\alpha,v)
+
{\cal Q}(\qs)-{\cal Q}(v)
+
3\k\frac{\ep^{-2v}}{\sqrt{v}}
+
\n\frac{\ep^{\oh \qs}}{\sqrt{\qs}}
\Bigr)
+2c\,,
\end{equation}
with the parameter $v$ varying from $0$ to $\qs$. The latter is a solution to Eq.\eqref{q} in the interval $[0,1]$. The functions ${\cal L}^+$ and ${\cal E}^+$ are as defined in Appendix A. $c$ is a normalization constant. The tangent angle $\alpha$ can be expressed in terms of the parameter $v$ by using the force balance equation at $r=\rv$. The result is given by Eq.\eqref{alpha1}.

Since $\ell$ is an increasing function of $v$, increasing $\ell$ leads to a situation where the vertex reaches the position of the light quark. In this case the configuration looks like that of Figure \ref{cQQq}(m). It differs from the first only by the absence of the string stretched between the vertex and light quark so that the quark sits on top of the vertex. Hence the distance $\ell$ is expressed in terms of $v$ and $\alpha$ by the same formula as before, only for another parameter range, whereas the energy by 

\begin{equation}\label{EQQqm} 
E_{\QQq}=\g\sqrt{\s}
\Bigl(
2{\cal E}^+(\alpha,v)
+
\frac{1}{{\sqrt{v}}}\bigl(
3\k\ep^{-2v}
+
\n\ep^{\oh v}
\bigr)
\Bigr)
+2c\,.
\end{equation}
Clearly, it can be obtained from \eqref{EQQqs} by formally setting $\qs=v$. In this case, $\alpha$ is expressed in terms of $v$ by Eq.\eqref{alpham}, with $\bar v$ replaced by $v$. By construction, the tangent angle must be non-negative. This condition allows one to find a range for $v$. It is given by $[\qs,\Vz]$, where $\Vz$ is a solution to Eq.\eqref{v0} (with $\bar v$ replaced by $v$). The meaning of the upper bound is that $\alpha(\Vz)=0$. In other words, there is no cusp at the string endpoints at $x=0$. 

From the expression \eqref{EQQqs}, it follows that $\ell$ remains finite at $v=\Vz$. The question now is how to reach larger values of $\ell$.  The answer is to consider negative values of $\alpha$. In that case, the configuration profile becomes convex near $x=0$, as shown in Figure \ref{cQQq}(l). The desired result is obtained from the previous one by replacing ${\cal L}^+$ and ${\cal E}^+$ with ${\cal L}^-$ and ${\cal E}^-$. So, 

\begin{equation}\label{EQQql}
\ell=\frac{2}{\sqrt{\s}}
{\cal L}^-(\lambda,v)
\,,
\qquad
E_{\QQq}=\g\sqrt{\s}
\Bigl(
2{\cal E}^-(\lambda,v)
+
\frac{1}{\sqrt{v}}
\bigl(3\k\ep^{-2v}
+
\n\ep^{\oh v}
\bigr)
\Bigr)
+2c\,.
\end{equation}
Here the parameter $v$ goes from $\Vz$ to $\Vo$, where $\Vo$ is a solution to Eq.\eqref{v1-QQq} in the interval $[0,1]$. $\lambda$ is the function of $v$ given by Eq.\eqref{lambda}. The strings approach the soft wall as $\lambda\rightarrow 1$. This places the upper bound on $v$. 

A summary of the above discussion is as follows. The energy of the connected configuration as a function of the heavy quark separation is given in parametric form by the two piecewise functions $E_{\QQq}=E_{\QQq}(v)$ and $\ell=\ell(v)$. 

For future reference, we give a brief description of $E_{\QQq}(\ell)$ for small and large $\ell$. It behaves for $\ell\rightarrow 0$ as 

\begin{equation}\label{EQQq-small}
E_{\QQq}(\ell)=E_{\QQ}(\ell)+E_{\qQb}+o(\ell)\,,
\qquad\text{with}
\qquad 
E_{\QQ}=-\frac{\alpha_{\QQ}}{\ell}+c+\boldsymbol{\sigma}_{\QQ}\ell
\,.
\end{equation}	
$E_{\qQb}$ is equal to $E_{\Qqb}$ defined in \eqref{Qqb} and the coefficients are given by 

\begin{equation}\label{coeff}
	\alpha_{\QQq}=-l_0E_0\g\,,
	\qquad
	\boldsymbol{\sigma}_{\QQq}=\frac{1}{l_0}\Bigl(E_1+\frac{l_1}{l_0}E_0\Bigr)\g\s
	\,,
\end{equation}
where $l_0=\frac{1}{2}\xi^{-\frac{1}{2}}B\bigl(\xi^2;\tfrac{3}{4},\tfrac{1}{2}\bigr)$, $l_1=\frac{1}{2}\xi^{-\frac{3}{2}}
\bigl[ \bigl(2\xi+\frac{3}{4}\frac{\k-1}{\xi}\bigr)B\bigl(\xi^2;\tfrac{3}{4},-\tfrac{1}{2}\bigr)-B\bigl(\xi^2;\tfrac{5}{4},-\tfrac{1}{2}\bigr)\bigr]$, $E_0=1+3\k+\frac{1}{2}\xi^{\frac{1}{2}}B\bigl(\xi^2;-\tfrac{1}{4},\tfrac{1}{2}\bigr)$, and $E_1=\xi\,l_1-1-6\k+\frac{1}{2}\xi^{-\frac{1}{2}}B\bigl(\xi^2;\tfrac{1}{4},\tfrac{1}{2}\bigr)$. Here $B(z;a,b)$ is the incomplete beta function and $\xi=\frac{\sqrt{3}}{2}(1-2\k-3\k^2)^{\frac{1}{2}}$. Thus the model we are considering has the desired property of factorization, expected from heavy quark-diquark symmetry \cite{wise}. 

$E_{\QQq}$ behaves for $\ell\rightarrow\infty$ as 

\begin{equation}\label{EQQq-large}
	E_{\QQq}=\sigma\ell-2\g\sqrt{\s}I_{\QQq}+2c+o(1)
	\,,
\qquad
\text{with} 
\qquad
	I_{\QQq}={\cal I}(\Vo)
	-
\frac{3\k\ep^{-2 \Vo}++\n\ep^{\oh\Vo}}{2\sqrt{\Vo}}
\,
\end{equation}
and the same string tension $\sigma$ as in \eqref{Eqqb}. The function ${\cal I}$ is defined in Appendix A.

A five-dimensional counterpart of the disconnected configuration of Figure \ref{4QQq}(b) is shown schematically in Figure \ref{con-ab}(b). It describes the non-interacting hadrons and, therefore, the total energy is the sum of the rest energies of the hadrons. Explicitly, it is given by Eq.\eqref{Eb}.  

Like in lattice QCD, the potential $V_0$ is given by the smallest eigenvalue of a model Hamiltonian 

\begin{equation}\label{HD-QQq}
{\cal H}(\ell)=
\begin{pmatrix}
E_{\QQq}(\ell) & \Theta_{\QQq} \\
\Theta_{\QQq} & E_{\Qqq}+E_{\Qqb} \\
\end{pmatrix}
\,,
\end{equation}
with $\Theta_{\QQq}$ describing the mixing between the two states. Explicitly, 

\begin{equation}\label{V0QQq}
V_0=\oh\Bigl(E_{\QQq}+E_{\Qqq}+E_{\Qqb}\Bigr)
-
\sqrt{\frac{1}{4}\Bigl(E_{\QQq}-E_{\Qqq}-E_{\Qqb}\Bigr)^2+\Theta_{\QQq}^2}
\,.	
\end{equation}

\begin{figure}[htbp]
\centering
\includegraphics[width=8.3cm]{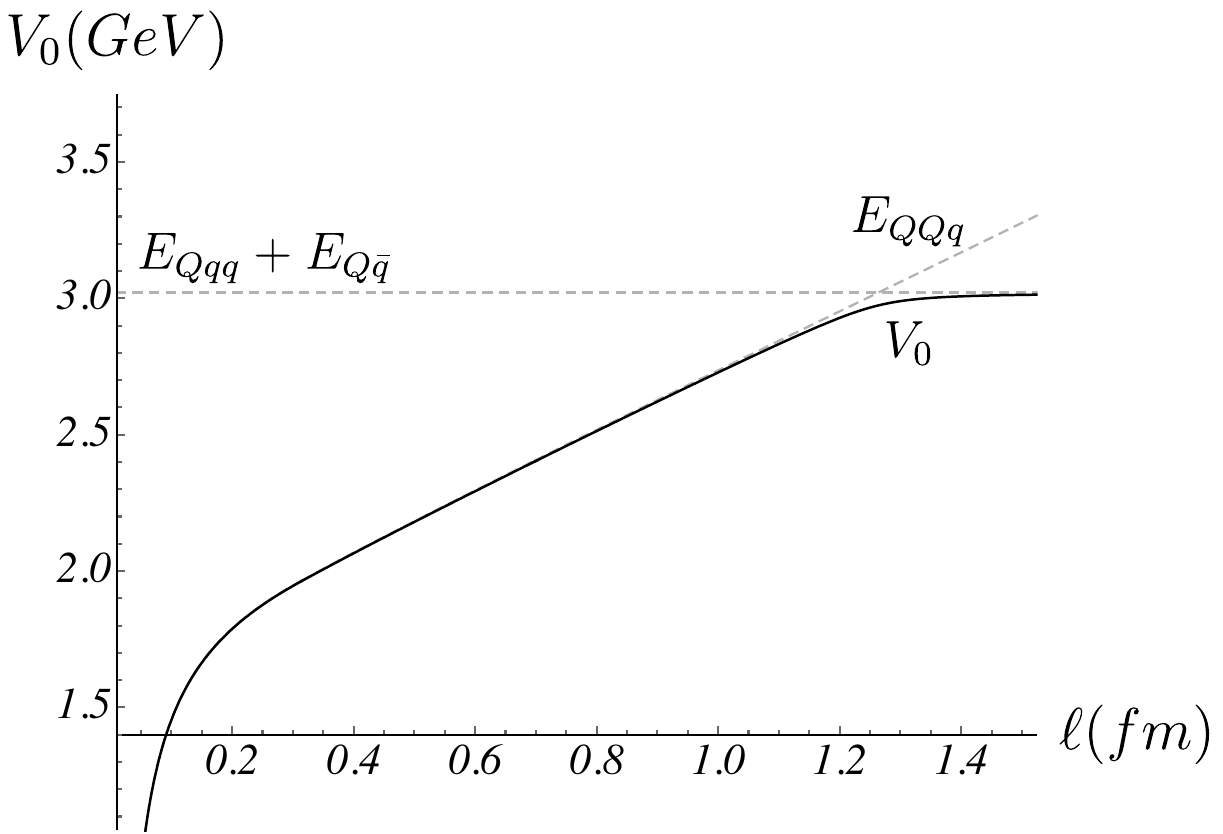}
\caption{{\small The potential $V_0$ of the $QQq$ system. Here $\Theta_{\QQq}=47\,\text{MeV}$.}}
\label{VQQq}
\end{figure}

It is instructive to give an example of this potential. For the same parameter values as in Figure \ref{Plotc}, it is shown in Figure \ref{VQQq}.
The potential asymptotically approaches $E_{\QQq}$ as $\ell$ tends to zero and $E_{\Qqq}+E_{\Qqb}$ as $\ell$ tends to infinity. The transition between these two regimes occurs around $\ell=\ell_{\QQq}$ which is a solution to the equation

\begin{equation}\label{lc-QQq}
E_{\QQq}(\ell_{\QQq})=E_{\Qqq}+E_{\Qqb} 
\,.	
\end{equation}
This equation simplifies drastically at large $\ell$, where the phenomenon of string breaking is expected to occur. Combining \eqref{EQQq-large} with \eqref{Eb}, one gets 

\begin{equation}\label{lcQQq}
\ell_{\QQq} =\frac{3}{\ep\sqrt{\s}}
\biggl(
{\cal Q}(\qs)-\frac{1}{3}{\cal Q}(\vs) 
+\k\frac{\ep^{-2\vs}}{\sqrt{\vs}}
+\n\frac{\ep^{\oh \qs}}{\sqrt{\qs}}
+\frac{2}{3}{\cal I}_{\QQq}
\biggr)
\,.
\end{equation}
For the parameter values of Sec.III, a simple estimate gives $\ell_{\QQq} =1.257\,\text{fm}$. 



\small
\begin{thebibliography}{99}

\bibitem{X38}
Belle Collaboration, S.-K. Choi {\it et al}., Phys. Rev. Lett. {\bf 91}, 262001 (2003). 
\bibitem{lebed}
R. F. Lebed {\it et al}., Summary of Topical Group on Hadron Spectroscopy (RF07) Rare Processes and Precision Frontier of Snowmass 2021, arXiv:2207.14594 [hep-ph].
\bibitem{book}
A. Ali, L. Maiani, and A.D. Polosa, Multiquark Hadrons, Cambridge University Press, 2019.
\bibitem{P43}
LHCb Collaboration, R. Aaij {\it et al.}, Phys.Rev.Lett. {\bf 122}, 222001 (2019).
\bibitem{bo}
M. Born and J.R. Oppenheimer, Annalen der Physik {\bf 389}, 457 (1927).
\bibitem{braat}
E. Braaten, C. Langmack, and D.H. Smith, Phys.Rev.D {\bf 90}, 014044 (2014). 
\bibitem{uaw}
J. Casalderrey-Solana, H. Liu, D. Mateos, K. Rajagopal, and U.A. Wiedemann, Gauge/String Duality, Hot QCD and Heavy Ion Collisions, Cambridge University Press, 2014.
\bibitem{a-QQq}
O. Andreev, J. High Energy Phys. {\bf 05} (2021) 173.
\bibitem{a-QQqq}
O. Andreev, Phys.Rev.D {\bf 105}, 086025 (2022).
\bibitem{a-QQbqqb}
O. Andreev, Phys.Rev.D {\bf 106}, 066002 (2022).
\bibitem{FK}
The well-understood example of such an analysis is the $Q\bar Q$ system. See, for example, the book by F. Knechtli, M. G\"unther and M. Peardon, Lattice Quantum Chromodynamics, Springer, 2017. 
\bibitem{XA}
X. Artru, Phys.Rept. {\bf 97}, 147 (1983); N. Isgur and J.E. Paton, Phys.Rev.D {\bf 31}, 2910 (1985).
\bibitem{XA2}
X. Artru, Nucl.Phys.B {\bf 85}, 442 (1975). 
\bibitem{a-strb}
O. Andreev, Phys.Lett.B {\bf 804} (2020) 135406; Phys.Rev.D {\bf 101}, 106003 (2020).
\bibitem{az1}
O. Andreev and V.I. Zakharov, Phys.Rev.D {\bf 74}, 025023 (2006).
\bibitem{white}
C.D. White, Phys.Lett.B {\bf 652}, 79 (2007).
\bibitem{a-3qPRD} 
 O. Andreev, Phys.Lett.B {\bf 756}, 6 (2016); Phys.Rev.D {\bf 93}, 105014 (2016).
\bibitem{witten}
E. Witten, J. High Energy Phys. {\bf 9807}, 006 (1998).
\bibitem{son}
J. Erlich, E. Katz, D.T. Son, and M.A. Stephanov, Phys.Rev.Lett. {\bf 95}, 261602 (2005).
\bibitem{bulava}
J. Bulava, B. H\"orz, F. Knechtli, V. Koch, G. Moir, C. Morningstar, and M. Peardon, Phys.Lett.B {\bf 793} (2019) 493.
\bibitem{a-stb3q}
O. Andreev, Phys.Rev.D {\bf 104}, 026005 (2021).
\bibitem{voloshin}
M.B. Voloshin, Deciphering the XYZ States, a talk at "{\it 17th Conference on Flavor Physics and CP Violation (FPCP 2019)}, arXiv:1905.13156 [hep-ph].
\bibitem{pion-factor}
P. Bicudo, A. Peters, S. Velten, and M. Wagner, Phys.Rev.D {\bf 103}, 114506 (2021); M. Sadl and S. Prelovsek, Phys.Rev.D {\bf 104}, 114503 (2021).
\bibitem{maiani}
L. Maiani, A.D. Polosa, and V. Riquer, Phys.Lett.B {\bf 749}, 289 (2015) 289. 
\bibitem{wolfram}
See, e.g., MathWorld - A Wolfram Web Resource. https://reference.wolfram.com/language/ref/ProductLog.html.
\bibitem{wise}
M.J. Savage and M.B. Wise, Phys.Lett.B {\bf 248}, 177 (1990).
\bibitem{a-q2}
O. Andreev, Phys.Rev.D {\bf 73}, 107901 (2006).
\bibitem{nucl}
L. Alvarez-Ruso, T. Ledwig, J. Martin Camalich, and M.J. Vicente-Vacas, Phys.Rev.D {\bf 88}, 054507 (2013). 
\bibitem{wagner}
P. Bicudo, K. Cichy, A. Peters, and M. Wagner, Phys.Rev.D {\bf 93}, 034501 (2016); P. Bicudo, M. Cardoso, A. Peters, M. Pflaumer, and M. Wagner, Phys.Rev.D {\bf 96}, 054510 (2017); M. Wagner, P. Bicudo, A. Peters, and S. Velten, Comparing meson-meson and diquark-antidiquark creation operators for a $\bar b\bar b ud$ tetraquark, a contribution to "The 38th International Symposium on Lattice Field Theory", LATTICE 2021, arXiv:2108.11731 [hep-lat].
\bibitem{quigg}
E.J. Eichten and C. Quigg, Phys.Rev.Lett. {\bf 119}, 202002 (2017); M. Karliner and J.L. Rosner, Phys.Rev.Lett. {\bf 119}, 202001 (2017).
\bibitem{wang}
Z.-G. Wang, Eur.Phys.J.C {\bf 78}, 826 (2018). 
\bibitem{FG}
F. Giannuzzi, Phys.Rev.D {\bf 99}, 094006 (2019). 
\bibitem{zhu}
R. Zhu, X. Liu, H. Huang, and C-F. Qiao, Phys.Lett.B {\bf 797}, 134869 (2019).  
\bibitem{lu}
Q.F. L\"u, D.Y. Chen, and Y.B. Dong, Phys.Rev.D {\bf 102}, 074021 (2020).
\bibitem{PDG}
P. A. Zyla {\it et al.} (Particle Data Group), PTEP {\bf 2020}, 083C01 (2020).

\end{thebibliography}
\end{document}